\documentclass[conference, onecolumn,12pt]{IEEEtran}
\usepackage{setspace}
\usepackage[margin=1in]{geometry}

\usepackage{amsthm}
\usepackage{amsmath, amssymb, amsfonts,bm}
\usepackage{graphicx}
\usepackage{epsfig}
\usepackage{epstopdf}
\usepackage{verbatim}
\usepackage{hyperref}
\usepackage{color}
\usepackage{caption}
\usepackage{subcaption}
\usepackage{cite}
\usepackage[none]{hyphenat}

\usepackage{tikz}
\usetikzlibrary{decorations.pathreplacing,positioning,shapes,shapes.geometric,calc,chains,shapes.multipart,arrows}
\tikzset{
	rubberduck/.style={
		shape=isosceles triangle,
		fill=red!30,
		minimum height=18pt,
		minimum width=12pt,
		shape border rotate=#1,
		isosceles triangle stretches,
		inner sep=0pt,
	},
	ap/.style={rubberduck=+90},
	queue/.style={
		rectangle split,
		minimum width=2em,
		rectangle split parts=4,
		draw,
		anchor=south,
	}
}

\newcommand{\Exp}{\mathsf E}

\newcommand{\Prob}{{\mathsf P}}
\newtheorem{theorem}{Theorem}
\newtheorem{lemma}{Lemma} 

\newtheorem{algorithm}{Algorithm}
\newtheorem{problem}{Problem}

\DeclareMathOperator{\subjectto}{subject\; to \;}

\DeclareMathOperator{\minimize}{minimize}

\IEEEoverridecommandlockouts

\begin{document}

\title{Dual-Timescale Spectrum Management in Small-Cell Wireless Networks}
\author{\IEEEauthorblockN{Fei Teng and Dongning Guo}
\IEEEauthorblockA{Department of Electrical Engineering and Computer Science\\ Northwestern University\\ Evanston, IL 60208
 \\  \\ \Large{January, 2019}
}}

\maketitle
\thispagestyle{plain}
\pagestyle{plain}

    \begin{abstract}

    To attain the targeted data rates of next generation cellular networks requires dense deployment of small cells in addition to macro cells which provide wide coverage.
    Dynamic radio resource management is crucial to the success of such heterogeneous networks due to much more pronounced traffic and interference variations in small cells.
    This work proposes a framework for spectrum management organized according to two timescales, which include 1) centralized optimization on a moderate timescale corresponding to typical durations of user sessions (several seconds to minutes in today's networks), and 2) distributed spectrum allocation on a fast timescale corresponding to typical latency requirements (a few milliseconds).
    An optimization problem is formulated to allocate resources on the slower timescale with consideration of (distributed) opportunistic scheduling on the faster timescale. Both fixed and fully flexible user association schemes are considered.
    Iterative algorithms are developed to solve these optimization problems efficiently for a cluster of cells with guaranteed convergence.
    Simulation results demonstrate advantages of the proposed framework and algorithms.

    \end{abstract}

    \section{Introduction}
    Dense deployment of small cells is a promising means to address the scarcity of spectrum resources for
    next-generation wireless networks~\cite{Cavalcanti-2005}. By reducing the coverage of each access point (AP) and increasing the density of APs, spectrum is reused more aggressively so that more links can transmit data simultaneously at higher rates, leading to much higher area spectral efficiency (in bits/second/m$^2$).

    In existing networks, radio resource management (RRM) is usually based either on full-spectrum reuse or fractional frequency reuse (FFR). In FFR, a main portion of the spectrum is reused everywhere except at cell edge, and the remaining spectrum is divided for orthogonal reuse at cell edge to avoid inter-cell interference~\cite{Lei-2007}.
    The success of existing RRM techniques depends on user and interference averaging in large cells. Such techniques are not as efficient in small cells due to more pronounced traffic and interference variabilities across cells.

    To address highly dynamic traffic in small cells, we believe it is best to organize resource allocation according to traffic, channel, and interference conditions on two different timescales.
    On a moderate timescale, a central controller allocates resources across many (e.g., hundreds of) cells periodically according to the anticipated traffic distribution and channel
    conditions in the next period. On a faster timescale, each cell further schedules links within the cell onto the cell's allocated resources from the moderate timescale.
    Due to latency requirements (often in milliseconds), the fast-timescale allocation is most likely carried out in a distributed manner
    based on instantaneous local traffic and channel conditions of each AP or within each small neighborhood of a few APs. In contrast, the moderate-timescale updates should allow collection of traffic and channel conditions across a large region and enough computation time. In the meantime, the moderate-timescale updates should be sufficiently frequent to track the aggregate traffic volume of user sessions in each small area. The moderate timescale is conceived to be in seconds or a few minutes in 5G networks.

    We introduce resource allocation with both fixed and flexible user association schemes.
    Assuming no fast-timescale information exchange between cells,~\cite{ZB-2014} formulated a traffic-driven spectrum allocation problem on a moderate timescale, where full-spectrum reuse and FFR are special cases.
    The treatment was generalized in \cite{ZB-2018} to also incorporate optimized user association and cell activation and implemented using a distributed approach.
    A scalable reformulation was developed and efficient algorithms were proposed to obtain near-optimal allocations for a network with up to 1000 APs in \cite{ZZ-2017}.
    Given the moderate-timescale traffic intensities of all user devices that need to be supported, the controller basically seeks the optimal set of {\em transmit patterns} across the entire network. Precisely, a {\em pattern} corresponds to a subset of APs. The notion of pattern finds its root in the idea of independent set scheduling in networks described by a conflict graph.
    Under mild assumptions, the problem of allocating resources to patterns is shown to be convex. 
    References~\cite{ZB-2014, ZB-2017, ZZ-2017} assume the moderate-timescale allocation is entirely in the frequency domain so that it is performed separately from the fast-timescale allocation.

    The goal of the current work is to study interactions between the fast- and moderate-timescale allocations.
    In addition to global information collection by a central controller on the moderate timescale, we allow instantaneous fast-timescale information exchange between nearby APs. This enables opportunistic scheduling, where an AP may use time resources not allocated to it when no nearby AP has data to send. The formulation here accounts for interactions between the queues of different devices due to interference. Since there is no known expression for the delay of such interactive queueing systems, an approximation is obtained by using
    the AP utilizations as a surrogate for the amount of interactions.
    Assuming fixed user association, \cite{Teng-2015} formulated a mixed-timescale spectrum allocation problem using simple M/M/1-queue approximations.
    This work considers both fixed and flexible user association schemes and uses M/G/1-queue approximations. We introduces three formulations with increasing capabilities and complexities. 
    The final optimization problem is essentially bi-convex, so we propose iterative algorithms to solve it with relatively low computational complexity and provable convergence.
    Simulation results demonstrate substantial delay reduction compared to the schemes in~\cite{ZB-2014}.

    This work on dual-timescale spectrum and time allocation is unique in the literature. Most work, such as \cite{Chang-2009, Ali-2009, Madan-2010, Liao-2014}, considered spectrum allocation on a single timescale.
    The authors of \cite{Rengarajan-2008} devised scheduling policies in the
    time domain, but considered full-spectrum reuse only.
    As in \cite{Lim-2014, Shen-2014, Deb-2014}, the spectrum allocation problem is often formulated as a discrete optimization problem (we study a continuous one here).
    Joint user association and spectrum allocation is explored here because of their coupled nature \cite{ZB-2018,Ye-2013,Fooladivanda-2013,Kuang-2014, Kuang-2016, Dotzler-2010}.
    In particular, by assuming that every cell uses all the spectrum, the authors of \cite{Ye-2013} investigated a joint user association and intra-cell resource allocation problem.
    In \cite{Fooladivanda-2013}, joint multi-cell channel allocation and user association was studied. However, \cite{Fooladivanda-2013} only allowed three pre-defined resource allocation strategies, namely, orthogonal deployment, co-channel deployment, and partially shared deployment. This paper considers more flexible spectrum allocation and user association than all the preceding work. Although references \cite{ZB-2018, Kuang-2014, Dotzler-2010} also considered flexible resource allocation schemes, the treatments therein were for a single timescale.

    The rest of this paper is organized as follows.
    The system model is introduced in Section~\ref{Ch:RRM SE:System model}.  Three increasingly more complex problem formulations are presented in Sections~\ref{Spectrum allocation for fixed user association}, \ref{Spectrum allocation with scheduling for fixed user association}, and~\ref{Spectrum allocation with scheduling for flexible user association}, respectively.  Numerical results are given in Section~\ref{Simulation results}. Concluding remarks are given in Section~\ref{Conclusion}.


    \section{System Model} \label{Ch:RRM SE:System model}

    We consider the downlink of a wireless heterogeneous network with $n$ access points, including possibly macro, pico, and other small cell base transceiver stations.
    Denote the set of all AP indices as $\mathcal{N}=\{1,\ldots,n\}$. Two APs are said to be neighbors of each other if they can fully exchange their
    traffic and channel state information with negligible latency.

    We denote the set of all device indices as $\mathcal{K}=\{1,\ldots,k\}$.
    A packet here models a user session, which is typically much longer than a network-layer datagram. The moderate-timescale traffic of device $j$ is modeled by a homogeneous Poisson point process with arrival rate $\lambda_j$ packets per second. The length of each packet is an independent exponentially distributed random variable with mean $L$ (bits).

    Suppose APs operate on a (licensed) frequency band of $W$ Hz in total. 
    The frequency resources are assumed to be homogeneous on the moderate timescale. A transmit pattern is simply a subset of APs.  A certain resource in time or frequency or both is said to be allocated to a pattern if the resource is to be shared by APs in that pattern but no other APs. Spectrum allocation to the $n$ APs can be viewed as a division of the spectrum into $2^n$ segments corresponding to all possible patterns. It suffices to describe an partition of the spectrum as $\{y^\mathcal{F}\}_{\mathcal{F}\subset \mathcal{N}}$, where $y^\mathcal{F}$ is the fraction of total bandwidth allocated to pattern $\mathcal{F}\subset \mathcal{N}$. We have
    \begin{align} \label{eq:total_one}
    \sum_{\mathcal{F}\subset \mathcal{N}} y^\mathcal{F} =1.
    \end{align}
    Any efficient allocation should set the bandwidth of the empty pattern to zero, i.e., $y^\phi=0$.

\begin{figure}
    	\centering
      \begin{tikzpicture}[x=1cm,y=1cm,node distance=0 cm,outer sep=0pt]
        \tiny
        \tikzstyle{specb}=[rectangle,draw,anchor=south west,text centered,minimum height=7mm,fill=blue!50]
        \tikzstyle{specg}=[rectangle,draw,anchor=south west,text centered,minimum height=7mm,fill=green!50]
        \tikzstyle{spec}=[rectangle,draw,anchor=south west,text centered,minimum height=7mm,fill=gray!50]

        \draw[color=white] (0,0) -- (0,2.1) -- (7.6,2.1) -- (7.6,0) -- (0,0);

    \draw (0,0) -- (0,2.1) -- (7.6,2.1) -- (7.6,0) -- (0,0);
    \node at (-0.4,0.3) {\small AP 3};
    \node at (-0.4,1) {\small AP 2};
    \node at (-0.4,1.7) {\small AP 1};
    \node at (4,-.75) {\small frequency/time resources};

    \node[spec,minimum width=6mm] (1_1) at (0,1.4) {};

    \draw[decorate, decoration={brace,mirror}, blue](0.025,-.03)--(.575,-.03);
    \node at (0.3,-.4) {\small $y^{\{1\}}$};

    \node[spec,minimum width=10mm] (2_2) [below right = of 1_1]{};

    \draw[decorate, decoration={brace,mirror}, blue] (.625,-.03) -- (1.575,-.03);
    \node at (1.1,-.4) {\small $y^{\{2\}}$};

    \node[spec,minimum width=6mm] (3_3) [below right = of 2_2]{};

    \draw[decorate, decoration={brace,mirror}, blue] (1.625,-.03) -- (2.175,-.03);
    \node at (1.8,-.4) {\small $y^{\{3\}}$};

    \node[spec,minimum width=9mm] (12_1) at (2.2,1.4) {};
    \node[spec,minimum width=9mm] (12_2) [above right = of 3_3]{};

    \draw[decorate, decoration={brace,mirror}, blue] (2.225,-.03) -- (3.075,-.03);
    \node at (2.7,-.4) {\small $y^{\{1,2\}}$};

    \node[spec,minimum width=11mm] (23_2) [right = of 12_2]{};
    \node[spec,minimum width=11mm] (23_3) [below right = of 12_2]{};

    \draw[decorate, decoration={brace,mirror}, blue] (3.125,-.03) -- (4.175,-.03);
    \node at (3.7,-.4) {\small $y^{\{2,3\}}$};

    \node[spec,minimum width=21mm] (13_1) [above right = of 23_2]{};
    \node[spec,minimum width=21mm] (13_3) [below right = of 23_2]{};

    \draw[decorate, decoration={brace,mirror}, blue] (4.225,-.03) -- (6.275,-.03);
    \node at (5.2,-.4) {\small $y^{\{1,3\}}$};

    \node[spec,minimum width=13mm] (123_1) [right = of 13_1]{};
    \node[spec,minimum width=13mm] (123_2) [below right = of 13_1]{};
    \node[spec,minimum width=13mm] (123_3) [right = of 13_3]{};

    \draw[decorate, decoration={brace,mirror}, blue] (6.325,-.03) -- (7.575,-.03);
    \node at (6.8,-.4) {\small $y^{\{1,2,3\}}$};

    \node[specb,minimum width=3mm] (1_1-1) at (0,1.4) {1};
    \node[specg,minimum width=3mm] (1_1-2) [right = of 1_1-1]{2};
    \node[specb,minimum width=5mm] (2_2-1) [below right = of 1_1-2]{1};
    \node[specg,minimum width=5mm] (2_2-2) [right = of 2_2-1]{2};
    \node[specb,minimum width=3mm] (3_3-1) [below right = of 2_2-2]{1};
    \node[specg,minimum width=3mm] (3_3-2) [right = of 3_3-1]{2};

    \node[specb,minimum width=6mm] (12_1-1) at (2.2,1.4) {1}; 
    \node[specg,minimum width=3mm] (12_1-2) [right = of 12_1-1]{2};
    \node[specb,minimum width=3mm] (12_2-1) [above right = of 3_3-2]{1};
    \node[specg,minimum width=6mm] (12_2-2) [right = of 12_2-1]{2};
    \node[specb,minimum width=6mm] (23_2-1) [right = of 12_2-2]{1};
    \node[specg,minimum width=5mm] (23_2-2) [right = of 23_2-1]{2};
    \node[specb,minimum width=5mm] (23_3-1) [below right = of 12_2-2]{1};
    \node[specg,minimum width=6mm] (23_3-2) [right = of 23_3-1]{2};
    \node[specb,minimum width=9mm] (13_1-1) [above right = of 23_2-2]{1};
    \node[specg,minimum width=12mm] (13_1-2) [right = of 13_1-1]{2};
    \node[specb,minimum width=12mm] (13_3-1) [below right = of 23_2-2]{1};
    \node[specg,minimum width=9mm] (13_3-2) [right = of 13_3-1]{2};

    \node[specb,minimum width=5mm] (123_1-1) [right = of 13_1-2]{1};
    \node[specg,minimum width=8mm] (123_1-2) [right = of 123_1-1]{2};
    \node[specb,minimum width=9mm] (123_2-1) [below right = of 13_1-2]{1};
    \node[specg,minimum width=4mm] (123_2-2) [right = of 123_2-1]{2};
    \node[specb,minimum width=6mm] (123_3-1) [right = of 13_3-2]{1};
    \node[specg,minimum width=7mm] (123_3-2) [right = of 123_3-1]{2};

    \draw[decorate, decoration={brace}, blue] (4.225,2.12) -- (5.075,2.12);
    \draw[decorate, decoration={brace}, blue] (5.125,2.12) -- (6.275,2.12);
    \draw[decorate, decoration={brace}, blue] (4.225,0.72) -- (5.375,.72);
    \draw[decorate, decoration={brace}, blue] (5.425,0.72) -- (6.275,.72);
    \node at (4.6,2.5) {\small $x_{1\rightarrow 1}^{\{1,3\}}$};
    \node at (5.7,2.5) {\small $x_{1\rightarrow 2}^{\{1,3\}}$};
    \node at (4.8,1.05) {\small $x_{3\rightarrow 1}^{\{1,3\}}$};
    \node at (5.9,1.05) {\small $x_{3\rightarrow 2}^{\{1,3\}}$};

    \draw (.5,5) node[ap] (apone) {\small AP1};
    \draw (3.5,5) node[ap] (aptwo) {\small AP2};
    \draw (6.5,5) node[ap] (apthree) {\small AP3};
    \node[draw,fill=blue!50] (ueone) at (2,3.5) {\small device1};
    \node[draw,fill=green!50] (uetwo) at (5,3.5) {\small device2};
    \draw [->] (apone) -- (ueone);
    \draw [->] (aptwo) -- (ueone);
    \draw [->] (aptwo) -- (uetwo);
    \draw [->] (apthree) -- (uetwo);
    \draw [->] (apthree) -- (ueone);
    \draw [->] (apone) -- (uetwo);
    \end{tikzpicture}
    	\caption{Illustration of all patterns of a 3-AP 2-device network with spectrum allocation variables.}
    \label{fig: spectrum_allocation}
    \end{figure}
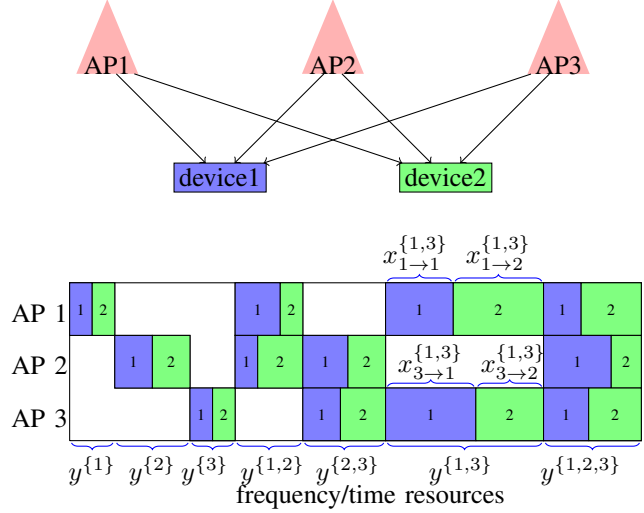

    Each AP may serve a set of devices, and each device may be served by any allocated subset of APs using any time and spectrum resources.
    To be specific, denote the fraction of total bandwidth used by AP $i$ to serve device $j$ over frequency pattern $\mathcal{F}$ as $x_{i\rightarrow j}^\mathcal{F}$, which is only defined for $i\in \mathcal{F}$.
    Throughout this paper, we assume APs do not support broadcast coding schemes.
    That is, if an AP transmits data to multiple devices simultaneously, those transmissions must be over nonoverlapping subsets of the spectrum.
    Then, for every $\mathcal{F}\subset \mathcal{N}$ and $i\in \mathcal{F}$,
    \begin{align}
    \sum_{j \in \mathcal{K}}x_{i\rightarrow j}^\mathcal{F} \leq y^\mathcal{F}. \label{eq:relate_x_y}
    \end{align}
    An example of spectrum allocation for three APs is shown in Fig. \ref{fig: spectrum_allocation}, where $y^{\{1,3\}}$ is the fraction of total bandwidth allocated to pattern $\{1,3\}$. AP 1 assigns $x_{1\rightarrow 1}^{\{1,3\}} \cdot W$ Hz to device 1 and $x_{1\rightarrow 2}^{\{1,3\}} \cdot W$ Hz to device 2, while AP 3 assigns $x_{3\rightarrow 1}^{\{1,3\}} \cdot W$ Hz to device 1 and $x_{3\rightarrow 2}^{\{1,3\}} \cdot W$ Hz to device 2.

    One unique feature in this paper is to consider allocation of time resources to patterns in a similar fashion as spectrum allocation.
    The patterns $\{\mathcal{T}\subset \mathcal{N}\}$ can be employed in a round robin fashion periodically. For convenience, we refer to one period as one time unit.
    Let $z^\mathcal{T}$ be the fraction of a time unit allocated to time pattern $\mathcal{T}$. Clearly, $\sum_{\mathcal{T}\subset \mathcal{N}} z^\mathcal{T} =1$.
    Here we assume a time unit is small relative to the moderate timescale, so that we can ignore the relative scheduling delay of different time patterns.

    In general, the time allocation can be different for different spectrum segments.  To prevent the treatment from becoming unwieldy, we assume identical time allocation for all spectrum segments. As a result, we simply divide all resources into orthogonal rectangular frequency-time physical resource blocks (PRBs) in similar manner as in Long Term Evolution (LTE) standards.
    Each PRB is indexed by $(\mathcal{F},\mathcal{T})$, where $\mathcal{F}$ is the
    pattern in frequency and $\mathcal{T}$ the pattern in time. The bandwidth of the PRB is $W\cdot y^\mathcal{F}$ Hz and the duration is $z^\mathcal{T}$ time units.

    If resource allocation occurs only on the moderate timescale, it is apparently inconsequential whether the allocation is in time, in frequency, or in both.  However, the actual allocation can be further adjusted on a fast timescale depending on instantaneous traffic.
    On a fast timescale, an AP may or may not have data to send at a given point in time. If an AP does not transmit over a given resource, it is said to be \emph{silent} over the resource.

    We consider a simple and fair scheduling rule:
    AP $i$ may transmit on a PRB indexed by $(\mathcal{F}, \mathcal{T})$ only if the following conditions hold:
    \begin{itemize}
    \item AP $i$ belongs to the spectrum pattern $\mathcal{F}$, i.e., $i\in \mathcal{F}$;
    \item AP $i$ has data (for some device(s));
    \item Either AP $i$ also belongs to the time pattern $\mathcal{T}$, i.e., $i\in \mathcal{T}$, or all neighbors of AP $i$ are silent on this PRB.
    \end{itemize}
    The key here is that APs are allowed to opportunistically employ more resources when their respective neighbors are silent.

\begin{figure}
    \centering
    \includegraphics[angle=360,width=.5\columnwidth]{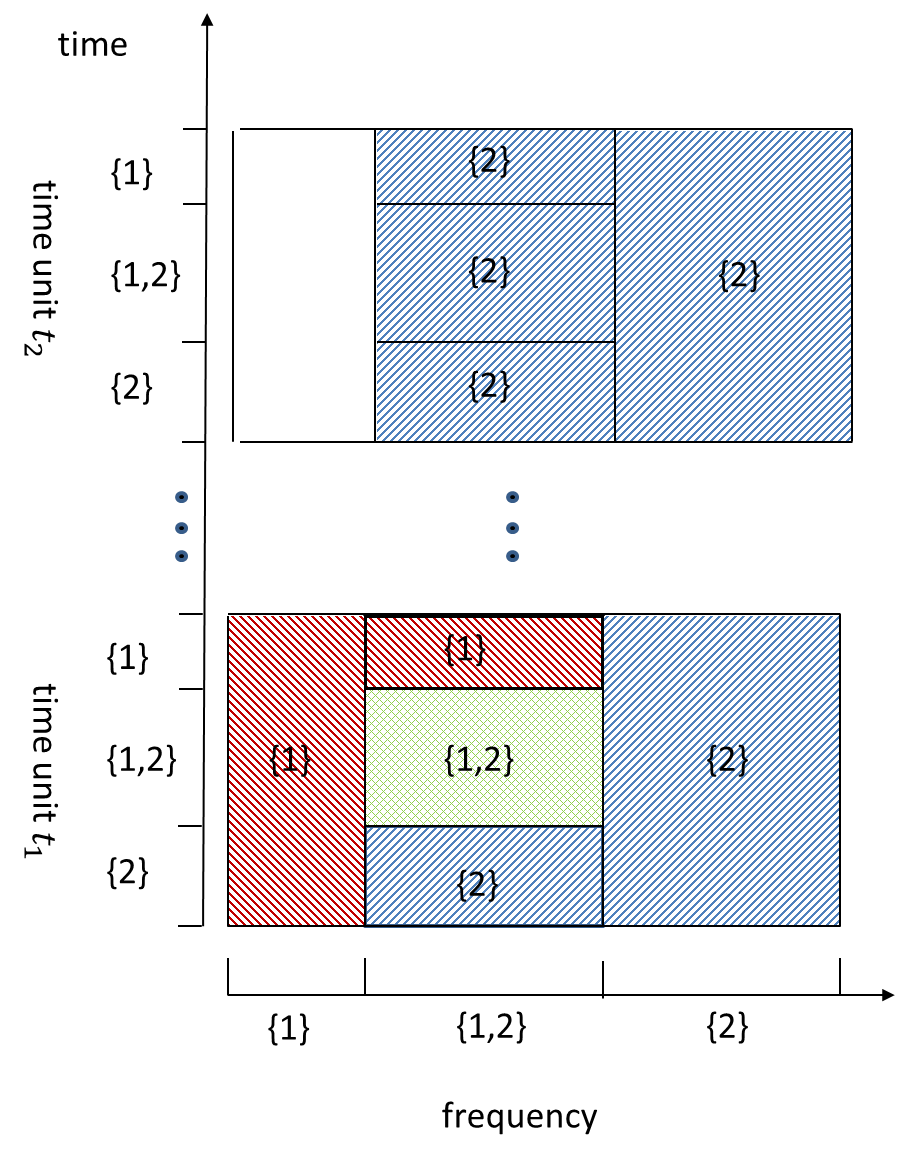}
    \caption{Frequency-time resource blocks for two neighboring APs: the set in the shadowed area represents the set of transmitting APs in each resource block.}
    \label{fig:time_freq_blocks}
    \end{figure}

    An example allocation to two neighboring APs is shown in Fig.~\ref{fig:time_freq_blocks}. The spectrum segments $\{1\}$ and $\{2\}$ are used exclusively by AP 1 and AP 2, respectively. Since those segments are for exclusive use, there is no need to divide them in the time domain. The spectrum segment $\{1,2\}$ is shared between the two APs, which is further divided into three time patterns: $\{1\}$, $\{2\}$, and $\{1,2\}$. These time resources can be reallocated opportunistically depending on instantaneous traffic conditions. Specifically, the frequency-time resource allocation at time unit $t_1$ in Fig.~\ref{fig:time_freq_blocks} corresponds to the case where both APs have data, and that at time unit $t_2$ in Fig.~\ref{fig:time_freq_blocks} corresponds to the case where only AP 2 has data. In the latter case, AP 2 may take all the spectrum shared by the two AP's, including the time patterns originally allocated to AP 1.

    Assume that AP $i$ transmits 
    with fixed flat transmit power spectral density (PSD) $q_i$.
    If multiple APs transmit to the same device over the same resource, then data from each AP may be decoded with all other transmissions treated as interference or data from all APs may be jointly using same multi-user decoding technique.
    Let $s_{i\rightarrow j}^\mathcal{A}$ denote the spectral efficiency of the link from AP $i$ to device $j$ under pattern $\mathcal{A}$ (i.e., when APs in $\mathcal{A}$ simultaneously transmit over the resource).
    For concreteness in obtaining numerical results in Section \ref{Simulation results}, we treat interference as noise and use Shannon's formula to obtain the spectral efficiency of the link from AP $i$ to device $j$ under pattern $\mathcal{A}$:
        \begin{align} \label{eq:SE}
        s_{i\rightarrow j}^\mathcal{A} = \log_2\left(1+ \frac{q_i h_{i \rightarrow j}}{\sum_{i'\in \mathcal{A}\setminus \{i\}}  q^{i'} h_{i' \rightarrow j} + n_j}\right) \ \text{bits/s/Hz}
        \end{align}
    when $i\in \mathcal{A}$, where $h_{i \rightarrow j}$ is the power gain of the link from AP $i$ to device $j$, and $n_j$ is the noise PSD at device $j$. If $i \not\in \mathcal{A}$, we set $s_{i\rightarrow j}^\mathcal{A} = 0$.
    On the moderate timescale, the link gain $h_{i \rightarrow j}$ is treated as constant and flat over the entire spectrum, which includes the effects of path loss and shadowing.
    Evidently, we have
    \begin{align} \label{eq:SE_comparison}
        s_{i\rightarrow j}^\mathcal{A}\geq s_{i\rightarrow j}^\mathcal{B},  \ \forall \mathcal{A}\subset \mathcal{B}\subset  \mathcal{N}, i\in \mathcal{A}, j\in \mathcal{K}.
    \end{align}
    It should become clear that we distinguish resources based on their patterns because the actual set of APs transmitting over a resource determines the quality of all links over the resource.

    If AP $i$ serves a single fixed device, we drop the device index $j$ and use $s^{\mathcal{A}}_i$ to represent the spectral efficiency of the link under pattern $\mathcal{A}$.
    For ease of comprehension, the device index $j$ has been used to denote an individual device so far. 
    We may follow~\cite{ZB-2014, ZB-2018} to reduce the model complexity by treating devices near each other with similar quality of service (QoS) requirements collectively as a user device group.
    In this case, resources are allocated to groups based on the average channel conditions and aggregate traffic conditions of the groups. Finer allocation is then carried out on a faster timescale.
    All the preceding notions of an individual device can be generalized to a user device group.


    \section{Slow-timescale Spectrum Allocation with Fixed User Association} \label{Spectrum allocation for fixed user association}

    We begin with moderate-timescale spectrum allocation with fixed user association.
    For simplicity, we assume that each AP serves a single device associated to it. 
    In a slot when the set of APs with data is $\mathcal{B}$, the rate contributed by frequency pattern $\mathcal{F}$ to AP $i$ is $s^{\mathcal{F}\cap \mathcal{B}}_i y^\mathcal{F} W$ bits/second, i.e., the product of the spectral efficiency and the pattern's bandwidth.
    The total service rate of AP $i$ under pattern $\mathcal{B}$ is thus
      \begin{align} \label{eq_rate_no_schedule}
        r^\mathcal{B}_i= \frac{W}{L} \sum_{\mathcal{F}\subset \mathcal{N}} s^{\mathcal{F}\cap \mathcal{B}}_i y^\mathcal{F} \ \ \text{packets/second},
      \end{align}
    where we include the factor $\frac{W}{L}$ to change the units to packets/second for later convenience.
     All statistics of the queueing system are determined by service rates $(r^\mathcal{B}_i)_{\mathcal{B}\subset \mathcal{N}, i\in \mathcal{B}}$ and arrival rates~$(\lambda_i)_{i\in \mathcal{N}}$.

    In a stable queueing system, AP $i$ has data to send over a fraction of time, which is referred to as the \emph{utilization} of AP $i$ for simplicity and devoted as $\rho_i$.
    It is important to note that $\rho_i$'s are consequences of the queueing dynamics.
    A stable queue must satisfy $0\leq \rho_i<1$.
    The analysis of general interactive queues is an open problem. In particular, there is no known explicit expression for the average delay in models with more than two interactive queues. To make progress, we make a simplifying assumption that the APs transmit independently despite the interactions.
    Then the probability that the set of APs with data is exactly $\mathcal{B}$ is
    \begin{align} \label{eq:prob_active_set}
    p^\mathcal{B}  = \left(\prod_{i\in \mathcal{B}} \rho_i\right) \left(\prod_{i\in \mathcal{N}\setminus \mathcal{B}}(1-\rho_i) \right).
    \end{align}
It is useful to note that
\begin{align}
  \sum_{\mathcal{B}\subset \mathcal{N}: i\in \mathcal{B}} p^\mathcal{B} = \rho_i.
\end{align}
Thus, $p^\mathcal{B}/\rho_i$ is a conditional probability, and is equal to the fraction of time that the pattern is $\mathcal{B}$ when $i\in \mathcal{B}$ has data.

     We develop an approximation for the average packet delay in the interactive queueing system.
    When AP $i$ transmits packets, its service rate is chosen from $2^{n-1}$ possible values which correspond to different sets of APs with data. Assuming that the service rate is constant during transmission of each packet, we use a M/G/1 queue with $2^{n-1}$ classes of packets to approximate the average delay. Each class corresponds to a specific subset of APs having data. In a stable queueing system, the number of packets served per second is equal to the arrival rate in steady state. Thus, we have
    \begin{align} \label{eq_balance}
     \sum_{\mathcal{B}\subset \mathcal{N}: i\in \mathcal{B}} p^\mathcal{B} r^\mathcal{B}_i = \lambda_i.
    \end{align}
    The average packet delay at AP $i$ can be derived according to the delay analysis for M/G/c queue with multiple customer classes \cite{Federgruen-1988}.
    \begin{lemma}\label{lemma:MG1_delay}
    The average delay of the M/G/1 queue is
    \begin{align} \label{eq:delay}
     d_i= \frac{\rho_i}{\lambda_i} + \frac1{1-\rho_i}\cdot\sum_{\mathcal{B}\subset \mathcal{N}: i\in \mathcal{B}} \frac{p^\mathcal{B}}{r^\mathcal{B}_i}, \ \forall i \in \mathcal{N} .
    \end{align}
    \end{lemma}
    Lemma \ref{lemma:MG1_delay} is proved in Appendix \ref{appdx:MG1_delay}.
    It is important to note that the right hand side of \eqref{eq:delay} is convex in $(r^\mathcal{B}_i)_{\mathcal{B}\subset \mathcal{N}}$ for fixed $(\rho_i)_{i\in \mathcal{N}}$ and $(p^\mathcal{B})_{\mathcal{B}\subset \mathcal{N}}$.
    Hereafter we let minimizing the approximate average delay \eqref{eq:delay} be the objective of resource allocation.

    \begin{problem}(Spectrum allocation on the moderate timescale) \label{Problem:fixed_UA_no_scheule}
      \begin{subequations}
       \begin{align}
        \underset{\bm{y},\bm{r}, \pmb{\rho}, \bm{p}, \bm{d}}{\minimize} \ &  \sum_{i=1}^n \lambda_i  d_i  \tag{P1a}\label{P1:obj} \\
        \subjectto
          &  d_i= \frac{\rho_i}{\lambda_i} + \frac1{1-\rho_i}\cdot \sum_{\mathcal{B}\subset \mathcal{N}: i\in \mathcal{B}} \frac{p^\mathcal{B}}{r^\mathcal{B}_i}, \ \forall i\in \mathcal{N} \tag{P1b}\label{P1:contr_b} \\
          &  r^\mathcal{B}_i=\frac{W}{L} \sum_{\mathcal{F}\subset \mathcal{N}} s^{\mathcal{F}\bigcap \mathcal{B}}_i y^\mathcal{F}, \ \forall \mathcal{B}\subset \mathcal{N}, \ i\in \mathcal{B} \tag{P1c}\label{P1:contr_c}  \\
          & \sum_{\mathcal{F}\subset \mathcal{N}}  y^\mathcal{F} =1 \tag{P1d}\label{P1:contr_e}\\
          & p^\mathcal{B}  = \left(\prod_{i\in \mathcal{B}} \rho_i\right) \left(\prod_{i\in \mathcal{N}\setminus \mathcal{B}}(1-\rho_i)\right), \ \forall  \mathcal{B}\subset \mathcal{N}  \tag{P1e}\label{P1:contr_f}\\
          &      \sum_{\mathcal{B}\subset \mathcal{N}: i\in \mathcal{B}} p^\mathcal{B} r^\mathcal{B}_i = \lambda_i , \ \forall i\in \mathcal{N} \tag{P1f}\label{P1:obj_relax} \\
          &  y^\mathcal{F} \geq 0, \ \forall \mathcal{F}\subset \mathcal{N}  \tag{P1g}\label{P1:contr_d} \\
          & 0\leq \rho_i < 1, \ \forall i\in \mathcal{N}. \tag{P1h} \label{P1:contr_h}
      \end{align}
      \end{subequations}
     \end{problem}

    The variables in Problem~\ref{Problem:fixed_UA_no_scheule} are $\bm{y}=(y^\mathcal{F})_{\mathcal{F}\subset \mathcal{N}},  \bm{r}=(r^\mathcal{B}_i)_{\mathcal{B}\subset \mathcal{N}, i\in \mathcal{B}}, \pmb{\rho}=(\rho_i)_{i\in \mathcal{N}}, \bm{p}=(p^\mathcal{B})_{\mathcal{B}\subset \mathcal{N}}$, and $\bm{d}=(d_i)_{i\in \mathcal{N}}$. The objective \eqref{P1:obj} divided by the total arrival rate $\sum_{i=1}^n \lambda_i$
    is equal to the (approximate) average packet delay of the entire network.
    Also, constraints \eqref{P1:contr_b}-\eqref{P1:obj_relax} are from~\eqref{eq:delay}, \eqref{eq_rate_no_schedule}, \eqref{eq:total_one}, \eqref{eq:prob_active_set}, and \eqref{eq_balance}, respectively. Constraint \eqref{P1:contr_d} assures all spectrum allocation to be nonnegative, and constraint \eqref{P1:contr_h} constrains that utilization of each queue must be between zero and one in a stable system.

    The throughput region of an interactive queueing system is given in \cite{Tassiulas-1922} in the context of resource allocation through coordinated scheduling in the time domain.
    When adapted to the spectrum allocation in \cite{ZB-2014}, the throughput region is expressed as (using the notation in this paper):
    \begin{align} \label{eq: lambda_region}
    \Lambda = \{(\lambda_1,\ldots, \lambda_n)| \exists (y^\mathcal{F})_{\mathcal{F}\subset \mathcal{N}} \text{ satisfying } \sum_{\mathcal{F}\subset \mathcal{N}}  y^\mathcal{F} =1 ,\text{ such that }     0\leq \lambda_i < \sum_{\mathcal{F}\subset \mathcal{N}} s^{\mathcal{F}}_i y^\mathcal{F}, \; \forall i\in \mathcal{N} \}.
    \end{align}
    In particular, $\Lambda$ is the convex hull of $\{(s^{\mathcal{F}}_i)_{i\in \mathcal{N}}, \mathcal{F}\subset \mathcal{N}\}$. 
    Problem~\ref{Problem:fixed_UA_no_scheule} is feasible for the entire throughput region in \eqref{eq: lambda_region}. Intuitively, if letting $\rho_i \simeq 1$ for all $i\in \mathcal{N}$ so that $p^\mathcal{B}$ is close to 1 if $\mathcal{B} = \mathcal{N}$ and 0 otherwise, we have $\lambda_i  \simeq r^\mathcal{N}_i = \sum_{\mathcal{F}\subset \mathcal{N}} s^{\mathcal{F}}_i y^\mathcal{F} $.


    Problem~\ref{Problem:fixed_UA_no_scheule} may be nonconvex due to the nonlinear equality constraints \eqref{P1:contr_f} and \eqref{P1:obj_relax}.
    However, we can use an iterative method to solve Problem~\ref{Problem:fixed_UA_no_scheule} with low complexity. We divide Problem~\ref{Problem:fixed_UA_no_scheule} into two sub-problems: The first one is to update $\bm{y}$ and $\bm{r}$ by fixing $\pmb{\rho}$ and $\bm{p}$; the second one is to update $\pmb{\rho}$ and $\bm{p}$ by fixing $\bm{y}$ and $\bm{r}$.  The basic idea for an efficient algorithm is to alternatively update the two groups of variables.

     \begin{problem}(For fixed $\pmb{\rho}$ and $\bm{p}$, update $\bm{y}$ and $\bm{r}$) \label{Problem:fixed_UA_no_scheule_sub1}
     \begin{subequations}
       \begin{align}
       \underset{\bm{y},\bm{r}}{ \minimize} \ &
        \sum_{i=1}^n \rho_i+ \sum_{\mathcal{B}\subset \mathcal{N}} p^\mathcal{B} \sum_{i\in \mathcal{B}} \frac{\lambda_i}{r^\mathcal{B}_i(1-\rho_i)} \tag{P2a}\label{SP1_obj} \\
        \subjectto
          &  r^\mathcal{B}_i=\frac{W}{L} \sum_{\mathcal{F}\subset \mathcal{N}} s^{\mathcal{F}\bigcap \mathcal{B}}_i y^\mathcal{F}, \ \forall \mathcal{B}\subset \mathcal{N}, \ i\in \mathcal{B}  \tag{P2b}\label{eq:r_constraint}\\
          &  y^\mathcal{F} \geq 0, \ \forall \mathcal{F}\subset \mathcal{N}   \tag{P2c}\\
          & \sum_{\mathcal{F}\subset \mathcal{N}}  y^\mathcal{F} =1,   \tag{P2d}\\
          & \sum_{\mathcal{B}\subset \mathcal{N}: i\in \mathcal{B}} p^\mathcal{B} r^\mathcal{B}_i = \lambda_i, \ \forall i\in \mathcal{N}. \tag{P2e}\label{ineq:decreasing_rho}
       \end{align}
     \end{subequations}
     \end{problem}
     The variables in Problem~\ref{Problem:fixed_UA_no_scheule_sub1} are $\bm{y}$ and $\bm{r}$.
     The objective \eqref{SP1_obj} is derived by substituting $d_i$ in \eqref{P1:obj} with \eqref{P1:contr_b}.
     Problem~\ref{Problem:fixed_UA_no_scheule_sub1} is a convex optimization problem because
     all constraints are linear and the objective is a linear combination of convex functions. Thus, this problem has a unique global minimum when feasible \cite{Bertsekas-1999}.

    Given $\bm{y}$ and $\bm{r}$, $\pmb{\rho}$ and $\bm{p}$ must satisfy \eqref{P1:contr_f} and \eqref{P1:contr_d}. The $2^n + n$ unknowns can be determined by the same number of equations. While we do not have closed form solution for $\pmb{\rho}$ and $\bm{p}$, we can obtain $\pmb{\rho}$ and $\bm{p}$ through an iterative algorithm. Here we use the method of interference function in \cite{Yates-1995} to update $\pmb{\rho}$ and $\bm{p}$.
    We define
    \begin{align}  \label{def:interf_func_0}
    g_i(\pmb{\rho})=\sum_{\mathcal{B}\subset \mathcal{N}: i\in \mathcal{B}} \left(\prod_{l\in \mathcal{B}\setminus \{i\}} \rho_l\right) \left(\prod_{l\in \mathcal{N}\setminus \mathcal{B}}(1-\rho_l) \right) r^\mathcal{B}_i.
    \end{align}
    Let $\bm{f}(\pmb{\rho})$ denote a vector of $n$ functions, where the $i$th function is defined as
    \begin{align}  \label{def:interf_func}
    f_i(\pmb{\rho})=\frac{\lambda_i}{g_i(\pmb{\rho})}.
    \end{align}
    By substituting \eqref{eq:prob_active_set} into \eqref{eq_balance}, we eliminate $(p^\mathcal{B})_{\mathcal{B}\subset \mathcal{N}}$ and obtain $\rho_i = f_i(\pmb{\rho})$, so $\pmb{\rho}$ satisfies the following fixed-point equation:
    \begin{align} \label{eq:rho}
    \pmb{\rho} = \bm{f}(\pmb{\rho}).
    \end{align}
    We obtain the following property of function $\bm{f}(\cdot)$ which is similar to that of the interference function in \cite{Yates-1995}.
    \begin{lemma} \label{lemma:properties}
    Suppose $(\lambda_i)_{i\in \mathcal{N}}$ and $(r^\mathcal{B}_i)_{\mathcal{B}\subset \mathcal{N}, i\in \mathcal{B}}$ are fixed. For every $i\in \mathcal{N}$, $f_i(\pmb{\rho})$ is monotonically increasing in every element of $\pmb{\widehat{\rho}}$. Equivalently, if $ 0\leq \pmb{\widehat{\rho}} \leq \pmb{\rho} < 1$, then $\bm{f}(\pmb{\widehat{\rho}}) \leq \bm{f}(\pmb{\rho}) $.
    \end{lemma}
    Lemma \ref{lemma:properties} is proved in Appendix~\ref{Appdx:lemma2}.
    For fixed $\bm{y}$ and $\bm{r}$, we update $\pmb{\rho}$ and $\bm{p}$ by using the following algorithm.
    \begin{algorithm}(For fixed $\bm{y}$ and $\bm{r}$, update $\pmb{\rho}$ and $\bm{p}$): \label{alg:update_rho}
    \begin{itemize}
        \item[1.] Initialization: set $\pmb{\rho}^{(0)}$ as current AP utilizations and let $\epsilon$ be a small positive constant;
        \item[2.] Update utilization and probabilities:
            Repeat
            \begin{align}
            & \pmb{\rho}^{(m+1)} = \bm{f}(\pmb{\rho}^{(m)})
            \end{align}
            until $|\pmb{\rho}^{(m)}-\pmb{\rho}^{(m-1)}| < \epsilon$;
        \item[3.] Let
            \begin{align}
                & \rho_i = \rho_i^{(m)}, \ \forall i\in \mathcal{N}
             \end{align}
             and compute $\left(p^\mathcal{B}\right)_{\mathcal{B}\in \mathcal{N}}$ by \eqref{eq:prob_active_set}.
     \end{itemize}
    \end{algorithm}

    \begin{lemma} \label{lemma:alg_rho_convg}
    Algorithm~\ref{alg:update_rho} converges as long as the initial point satisfies $\pmb{\rho}^{(0)}\geq \bm{f}(\pmb{\rho}^{(0)})$.
    \end{lemma}
    Lemma \ref{lemma:alg_rho_convg} is proved in Appendix~\ref{appdx:lemma4}. The key is to show that the iterative algorithm approaches the fixed point of \eqref{eq:rho}.

    Equipped with the preceding solution for the small problems,
    the proposed iterative algorithm for solving 
    Problem~\ref{Problem:fixed_UA_no_scheule} is described as follows:

    \begin{algorithm}(Iterative algorithm for solving Problem~\ref{Problem:fixed_UA_no_scheule}): \label{Alg:fixed_UA_no_scheule}
    \begin{itemize}
    \item[1.] Initialization: $\rho_i=1$, $\forall i\in \mathcal{N}$, $p^\mathcal{N}=1$ and $p^\mathcal{B}=0$ for all $\mathcal{B}\subset \mathcal{N}$ but $\mathcal{B} = \mathcal{N}$;
    \item[2.] Update $\bm{y}$ and $\bm{r}$ by solving Problem~\ref{Problem:fixed_UA_no_scheule_sub1};

    \item[3.] Update $\pmb{\rho}$ and $\bm{p}$ using Algorithm~\ref{alg:update_rho};
    \item[4.] Terminate if $\bm{y}$ has converged. Otherwise, return to step 2.

    \end{itemize}
    \end{algorithm}

    Algorithm~\ref{Alg:fixed_UA_no_scheule} mainly deals with the non-convexity nature of Problem~\ref{Problem:fixed_UA_no_scheule}.
    In part due to the monotonicity of $\bm{f}(\cdot)$,
    Algorithm~\ref{alg:update_rho} converges quite fast. Thus, Problem~\ref{Problem:fixed_UA_no_scheule} can be solved in an iterative way with manageable complexity as long as $n$ is not large.
    \begin{theorem} \label{Thm1:convergence}
    As long as $\pmb{\lambda}\in \Lambda$, Algorithm~\ref{Alg:fixed_UA_no_scheule} converges to a fixed point.
    \end{theorem}
    Theorem \ref{Thm1:convergence} is proved in Appendix~\ref{Appdx:thm1}.
    The solution derived through
    Algorithm~\ref{Alg:fixed_UA_no_scheule} may or may not be a local or global minimum.
    The effectiveness of the proposed framework and solution is validated through simulation in Section~\ref{Simulation results}.

    \section{Dual-Timescale Allocation with Fixed User Association} \label{Spectrum allocation with scheduling for fixed user association}

    In this section, we extend the model in Section \ref{Spectrum allocation for fixed user association} to describe a dual-timescale resource allocation problem.  While the resource allocation optimization is still carried out periodically on the same moderate timescale as before, the effect of opportunistic scheduling on the fast timescale is incorporated.
    It is assumed that, given the spectrum and time resources allocated on the moderate timescale, each cell exchanges instantaneous information about queue length and channel state with its neighboring cells and adjusts resource allocation accordingly on a fast timescale.

    As illustrated in Fig.~\ref{fig:time_freq_blocks}, time and frequency resources can be regarded as a collection of 2-dimensional blocks. During a given time slot, if the set of APs with data to send is $\mathcal{B}$, then the set of APs allowed to transmit under frequency pattern $\mathcal{F}$ is $\mathcal{F}\cap \mathcal{B}$. On PRB $(\mathcal{F},\mathcal{T})$, all APs in $\mathcal{F}\cap \mathcal{T}$ with data to send transmit, whereas the remaining ones in $\mathcal{F}\cap \mathcal{T}$ are silent.
    When some APs scheduled for PRB $(\mathcal{F},\mathcal{T})$ have no data to send, their selected neighbors take their places to transmit. This replacement selection can be itself a fast-timescale optimization problem. For concreteness, we introduce a specific simple selection method:
    As described in Section~\ref{Ch:RRM SE:System model}, if AP $i$ has data at time $\mathcal{T}$ and belongs to frequency pattern $\mathcal{F}$ (i.e., $i\in \mathcal{F}$), and all its neighboring APs are silent on PRB $(\mathcal{F},\mathcal{T})$,
    then AP $i$ uses PRB $(\mathcal{F},\mathcal{T})$ to transmit.
    The set of transmitting APs on PRB $(\mathcal{F},\mathcal{T})$ is determined by $(\mathcal{F},\mathcal{T},\mathcal{B})$ according to this rule. For the reader's convenience, definition of the sets $\mathcal{B}$, $\mathcal{F}$ and $\mathcal{T}$ are collected in Table \ref{table:sets}.

    \begin{table}[h!]
      \begin{center}
        \label{tab:table1}
        \begin{tabular}{l|l}
        \hline
          \textbf{Set} & \textbf{Set description} \\
        \hline
          $\mathcal{B}$ & the set of APs with data to transmit\\
          $\mathcal{F}$ & the set of APs allowed to transmit under frequency pattern $\mathcal{F}$ (strictly enforced) \\
          $\mathcal{T}$ & the set of APs that may transmit under time pattern $\mathcal{T}$ (not strictly enforced)\\
        \hline
        \end{tabular}
        \caption{Definitions of set of APs }
        \label{table:sets}
      \end{center}
    \end{table}

    In this section, we continue to assume that only one device is associated to each AP.
    Denote $\eta^{\mathcal{F},\mathcal{T},\mathcal{B}}_i$ as the spectral efficiency of AP $i$ on PRB $(\mathcal{F},\mathcal{T})$ when the set of APs having data is $\mathcal{B}$. In this case, suppose the actual set of transmitting APs on PRB $(\mathcal{F},\mathcal{T})$ as $\mathcal{A}$ according to the replacement selection method, then we have
    \begin{align}
    \eta^{\mathcal{F},\mathcal{T},\mathcal{B}}_i  & =s^\mathcal{A}_i, \label{ineq_SE_time_freq}
    \end{align}
    where the transmitting set $\mathcal{A}$ must be a subset of $\mathcal{B}\cap \mathcal{F}$.

    To analyze the average delay, consider the average service rate over one time unit. The rate of AP $i$ over PRB $(\mathcal{F},\mathcal{T})$ is $\eta^{\mathcal{F},\mathcal{T},\mathcal{B}}_i y^\mathcal{F} z^\mathcal{T}W$ bits/second. Hence, the average service rate of AP $i$ under pattern $\mathcal{B}$ is calculated as
    \begin{align} \label{eq_rate_with_schedule}
     r^\mathcal{B}_i= \frac{W}{L} \sum_{\mathcal{F}\subset \mathcal{N},\mathcal{T}\subset \mathcal{N}} \eta^{\mathcal{F},\mathcal{T},\mathcal{B}}_i y^\mathcal{F} z^\mathcal{T} \ \ \text{packets/second}.
    \end{align}

     \begin{problem}(Spectrum and time allocation with fixed user association) \label{Problem:fixed_UA_with_scheule}
      \begin{subequations}
       \begin{align}
        \underset{\bm{y}, \bm{z},\bm{r}, \pmb{\rho}, \bm{p}, \bm{d}}{\minimize} \ &  \sum_{i=1}^n \lambda_i d_i  \tag{P3a}\\
        \subjectto
          &  d_i= \frac{\rho_i}{\lambda_i} + \frac1{1-\rho_i}\sum_{\mathcal{B}\subset \mathcal{N}: i\in \mathcal{B}} \frac{p^\mathcal{B}}{r^\mathcal{B}_i}, \ \forall i\in \mathcal{N}  \tag{P3b} \\
          & r^\mathcal{B}_i= \frac{W}{L} \sum_{\mathcal{F}\subset \mathcal{N},\mathcal{T}\subset \mathcal{N}} \eta^{\mathcal{F},\mathcal{T},\mathcal{B}}_i y^\mathcal{F} z^\mathcal{T}, \ \forall \mathcal{B}\subset \mathcal{N}, \ i\in \mathcal{B}  \tag{P3c}\label{constr:p2_c}\\
          & \sum_{\mathcal{F}\subset \mathcal{N}}  y^\mathcal{F} =1,   \tag{P3d}\\
          & \sum_{\mathcal{T}\subset \mathcal{N}}  z^\mathcal{T} =1,   \tag{P3e}\label{constr:p2_2}\\
          &  z^\mathcal{T} \geq 0, \ \forall \mathcal{T}\subset \mathcal{N}   \tag{P3f}\label{constr:p2_1} \\
          &  y^\mathcal{F} \geq 0, \ \forall \mathcal{F}\subset \mathcal{N}   \tag{P3g} \\
          & 0\leq \rho_i < 1, \ \forall i\in \mathcal{N} \tag{P3h} \\
          & p^\mathcal{B}  = \left(\prod_{i\in \mathcal{B}} \rho_i\right) \left(\prod_{i\in \mathcal{N}\setminus \mathcal{B}}(1-\rho_i)\right), \ \forall  \mathcal{B}\subset \mathcal{N}  \tag{P3i} \\
          & \sum_{\mathcal{B}\subset \mathcal{N}: i\in \mathcal{B}} p^\mathcal{B} r^\mathcal{B}_i = \lambda_i, \ \forall i\in \mathcal{N}. \tag{P3j}
       \end{align}
      \end{subequations}
      \end{problem}

    The variables in Problem~\ref{Problem:fixed_UA_with_scheule} are $\bm{y},  \bm{z}, \bm{r},\pmb{\rho}, \bm{p}$ and $\bm{d}$. 
    In comparison with Problem~\ref{Problem:fixed_UA_no_scheule}, here we have additional variables $\bm{z}$ and corresponding constraints \eqref{constr:p2_2} and \eqref{constr:p2_1}. Because of time allocation, $r^\mathcal{B}_i$ in~\eqref{constr:p2_c} depends on both $\bm{y}$ and $\bm{z}$.
    The optimal delay of Problem~\ref{Problem:fixed_UA_with_scheule} is no greater than that of Problem~\ref{Problem:fixed_UA_no_scheule}. In fact by using the trivial time allocation of $z_\mathcal{N} =1$ and $z_\mathcal{T} = 0$ for all $ \mathcal{T}\subset \mathcal{N}$ in Problem~\ref{Problem:fixed_UA_with_scheule}, the problem degenerates to Problem~\ref{Problem:fixed_UA_no_scheule}.

    Similarly, Problem~\ref{Problem:fixed_UA_with_scheule} is in general nonconvex due to the nonlinear equality constraints. We can also use an iterative method to solve Problem~\ref{Problem:fixed_UA_with_scheule} as in Section~\ref{Spectrum allocation for fixed user association}. We divide Problem~\ref{Problem:fixed_UA_with_scheule} into three sub-problems: The first one is to update $\bm{y}$ and $\bm{r}$ by fixing $\pmb{\rho}, \ \bm{p}$ and $\bm{z}$; the second one is to update $\bm{z}$ and $\bm{r}$ by fixing $\pmb{\rho}, \ \bm{p}$ and $\bm{y}$; the third one is to update $\pmb{\rho}$ and $\bm{p}$ by fixing $\bm{y}$, $\bm{z}$ and~$\bm{r}$. 

    \begin{problem}(For fixed $\pmb{\rho}, \ \bm{p}$ and $\bm{z}$, update $\bm{y}, \ \bm{r}$) \label{Problem:fixed_UA_with_scheule_sub2}
      \begin{subequations}
       \begin{align}
        \underset{\bm{y}, \bm{r}}{\minimize} \ &
        \sum_{i=1}^n \rho_i+ \sum_{\mathcal{B}\subset \mathcal{N}} p^\mathcal{B} \sum_{i\in \mathcal{B}} \frac{\lambda_i}{r^\mathcal{B}_i(1-\rho_i)}  \tag{P4a} \\
        \subjectto
          & r^\mathcal{B}_i=\frac{W}{L} \sum_{\mathcal{F}\subset \mathcal{N}} \left(\sum_{\mathcal{T}\subset \mathcal{N}} \eta^{\mathcal{F},\mathcal{T},\mathcal{B}}_i z^\mathcal{T}\right) y^\mathcal{F}, \ \forall \mathcal{B}\subset \mathcal{N}, \ i\in \mathcal{B} \tag{P4b} \\
          &  y^\mathcal{F} \geq 0, \ \forall \mathcal{F}\subset \mathcal{N}  \tag{P4c} \\
          & \sum_{\mathcal{F}\subset \mathcal{N}}  y^\mathcal{F} =1,   \tag{P4d} \\
          & \sum_{\mathcal{B}\subset \mathcal{N}: i\in \mathcal{B}} p^\mathcal{B} r^\mathcal{B}_i = \lambda_i, \ \forall i\in \mathcal{N}. \tag{P4e}\label{ineq:decreasing_rho_1}
       \end{align}
      \end{subequations}
      \end{problem}
      Problem~\ref{Problem:fixed_UA_with_scheule_sub2} is identical to Problem~\ref{Problem:fixed_UA_no_scheule_sub1} with $s^{\mathcal{F}\bigcap \mathcal{B}}_i$ replaced by $\sum_{\mathcal{T}\subset \mathcal{N}} \eta^{\mathcal{F},\mathcal{T},\mathcal{B}}_i z^\mathcal{T}$.

     \begin{problem}(For fixed $\pmb{\rho}, \ \bm{p}$ and $\bm{y}$, update $\bm{z}, \ \bm{r}$) \label{Problem:fixed_UA_with_scheule_sub3}
      \begin{subequations}
       \begin{align}
        \underset{\bm{z}, \bm{r}}{\minimize} \ &
        \sum_{i=1}^n \rho_i+ \sum_{\mathcal{B}\subset \mathcal{N}} p^\mathcal{B} \sum_{i\in \mathcal{B}} \frac{\lambda_i}{r^\mathcal{B}_i(1-\rho_i)}  \tag{P5a} \\
        \subjectto
          & r^\mathcal{B}_i= \frac{W}{L} \sum_{\mathcal{T}\subset \mathcal{N}}  \left(z^\mathcal{T} \sum_{\mathcal{F}\subset \mathcal{N}} \eta^{\mathcal{F},\mathcal{T},\mathcal{B}}_i y^\mathcal{F}\right)  , \ \forall \mathcal{B}\subset \mathcal{N}, \ i\in \mathcal{B},  \tag{P5b} \\
          &  z^\mathcal{T} \geq 0, \ \forall \mathcal{T}\subset \mathcal{N}    \tag{P5c} \\
          & \sum_{\mathcal{T}\subset \mathcal{N}}  z^\mathcal{T} =1,   \tag{P5d} \\
          & \sum_{\mathcal{B}\subset \mathcal{N}: i\in \mathcal{B}} p^\mathcal{B} r^\mathcal{B}_i = \lambda_i, \ \forall i\in \mathcal{N}.  \tag{P5e}\label{ineq:decreasing_rho_2}
       \end{align}
      \end{subequations}
      \end{problem}

      Problems~\ref{Problem:fixed_UA_with_scheule_sub2} and~\ref{Problem:fixed_UA_with_scheule_sub3} are convex optimization problems. Thus, each problem has a unique global minimum when feasible.
     Given $\bm{y}$, $\bm{z}$ and $\bm{r}$, we update $\pmb{\rho}$ and $\bm{p}$ using Algorithm~\ref{alg:update_rho}. After introducing each update in the proposed method, we illustrate an iterative algorithm to solve Problem~\ref{Problem:fixed_UA_with_scheule_sub3}:

     \begin{algorithm}(Iterative algorithm for solving Problem~\ref{Problem:fixed_UA_with_scheule}): \label{Alg:fixed_UA_with_scheule}
    \begin{itemize}
    \item[1.] Initialization: $\rho_i=1$, $\forall i\in \mathcal{N}$; $p^\mathcal{B}=1$ if $\mathcal{B}=\mathcal{N}$, $p^\mathcal{B}=0$ otherwise;
    \item[2.] Update $\bm{y}, \ \bm{r}$ by solving Problem~\ref{Problem:fixed_UA_with_scheule_sub2};


    \item[3.] Update $\bm{z}, \ \bm{r}$ by solving Problem~\ref{Problem:fixed_UA_with_scheule_sub3};

    \item[4.] Update $\pmb{\rho}$, $\bm{p}$ by using Algorithm~\ref{alg:update_rho};
    \item[5.] Terminate if $\bm{y}$ has converged. Otherwise, return to step 2.

    \end{itemize}
    \end{algorithm}

    Problem~\ref{Problem:fixed_UA_with_scheule} can be solved in an iterative way with manageable complexity using Algorithm \ref{Alg:fixed_UA_with_scheule}.
    \begin{theorem} \label{Thm2:convergence}
     Algorithm~\ref{Alg:fixed_UA_with_scheule} converges to a fixed point.
    \end{theorem}
    The proof is similar to that for Theorem~\ref{Thm1:convergence}.
    For small networks, the solution derived through
    Algorithm~\ref{Alg:fixed_UA_with_scheule} may or may not be a local or global minimum.

    \section{Joint User Association and Dual-Timescale Allocation} \label{Spectrum allocation with scheduling for flexible user association}

    In this section, we allow fully flexible user association and complete the model for dual-timescale resource allocation.
    Each AP can serve multiple devices at the same time, and each device can be served simultaneously by multiple APs.
    Resources are now allocated to individual links.
    For simplicity, the time and frequency resources are still divided orthogonally as in Section \ref{Spectrum allocation with scheduling for fixed user association},
    where $z^{\mathcal{T}}$ denotes the fraction of time allocated to the time pattern $\mathcal{T}$,
    and $y^{\mathcal{F}}$ denotes the fraction of spectrum allocated to the frequency pattern $\mathcal{F}$.
    In addition, under frequency pattern $\mathcal{F}$, AP $i\in \mathcal{F}$ divides the spectrum $\left(y^{\mathcal{F}}W \text{ Hz }\right)$ to serve its associated devices,
    with $x_{i\rightarrow j}^\mathcal{F}W$ Hz allocated to serve device~$j$.
    Note that $x_{i\rightarrow j}^\mathcal{F}$ is only defined for $i\in \mathcal{F}$.
    If an AP serves multiple devices, it uses non-overlapping parts of the spectrum for different devices.
    Thus the allocation should satisfy~\eqref{eq:relate_x_y}.
    Given $\bm{x} = (x_{i\rightarrow j}^\mathcal{F})_{\mathcal{F}\subset \mathcal{N}, i\in \mathcal{N}, j\in \mathcal{K}}$, the set of APs transmitting data is determined by the set of devices receiving data $\mathcal{J}$ as
    \begin{align} \label{eq_B_by_J}
    \mathcal{B}(\bm{x},\mathcal{J}) = \{i: x_{i\rightarrow j}^\mathcal{F} >0, \
     \text{for some $j\in \mathcal{J}$ and $\mathcal{F}\subset \mathcal{N}$ with $i\in \mathcal{F}$}\}.
    \end{align}
    An example for two APs and two devices is as follows: Assume that $x_{i\rightarrow j}^{\mathcal{F}}> 0$ for every $\mathcal{F}\subset \{1,2\}, \ i\in \mathcal{F} \text{, and } j\in \{1,2\}$. From \eqref{eq_B_by_J} we have that $\mathcal{B}(\bm{x},\{1\}) = \{1,2\}$, which means when device 1 receives data, both APs 1 and 2 transit to device 1. Similarly, $\mathcal{B}(\bm{x}, \{2\}) = \{1,2\}$, and $\mathcal{B}(\bm{x}, \{1,2\}) = \{1,2\}$.  

    As in Section \ref{Spectrum allocation with scheduling for fixed user association}, if some APs scheduled for the frequency-time slot $(\mathcal{F},\mathcal{T})$ are silent on the PRB,
    their selected neighbors may take their places to transmit.
    The actual set of transmitting APs (denoted by $\mathcal{A}$) on the frequency-time slot $(\mathcal{F},\mathcal{T})$ is determined by $\mathcal{F}$, $\mathcal{T}$ and the set of APs having data $\mathcal{B}$.
    We denote the spectral efficiency on frequency-time slot $(\mathcal{F},\mathcal{T})$ from AP $i$ to device $j$ 
    when the set of APs having data is $\mathcal{B}$ as
    \begin{align}
    \eta^{\mathcal{F},\mathcal{T},\mathcal{B}}_{i\rightarrow j} =s^{\mathcal{A}}_{i\rightarrow j}.
    \end{align}
     The service rate of device $j$ contributed by AP $i$ on the frequency-time slot $(\mathcal{F},\mathcal{T})$ is then $ z^\mathcal{T}  \eta^{\mathcal{F},\mathcal{T},\mathcal{B}}_{i\rightarrow j} x_{i\rightarrow j}^\mathcal{F}W$ bits/second. 
    The total service rate of device~$j$ given the set of APs having data $\mathcal{B}$ is calculated as
    \begin{align} \label{eq_rate_with_schedule_UA_approx}
    r^\mathcal{B}_j=
     \frac{W}{L} \sum_{\mathcal{T}\subset \mathcal{N}}  z^\mathcal{T} \sum_{\mathcal{F}\subset \mathcal{N}} \left( \sum_{i\in \mathcal{F}}   \eta^{\mathcal{F},\mathcal{T},\mathcal{B}}_{i\rightarrow j} x_{i\rightarrow j}^\mathcal{F} \right)  \text{packets/second}.
    \end{align}

    All traffic for device $j$ arrives with a rate of $\lambda_j$ at a queue.
    Queues are interactive among devices.
    In a stable queueing system, device $j$ receives data over a fraction of time, which is referred to as the utilization of device $j$ and devoted as $\sigma_j$.
    Here we also assume that the time intervals that different devices receive are independent. We let $p^{\mathcal{B}}_j$ denote the conditional probability that APs in $\mathcal{B}$ have data given that device $j$ is receiving data. Given allocation $\bm{x}$, we then approximate the probability of the event that APs in $\mathcal{B}$ have data and device $j$ is receiving as
    \begin{align}
    \Prob\left(\text{APs in $\mathcal{B}$ have data, device $j$ is receiving}\right) = p^{\mathcal{B}}_j\times \sigma_j.
    \end{align}
    Because the set of receiving devices determines the set of APs with data according to \eqref{eq_B_by_J}, we can then write
    \begin{align} \label{eq:P_I}
     p^{\mathcal{B}}_j\times \sigma_j & = \sum_{\mathcal{J}\subset \mathcal{K}: j\in \mathcal{J}} \Prob\left(\text{APs in $\mathcal{B}$ have data, receivers in $\mathcal{J}$ are receiving}\right)   \\
     & =\sum_{\mathcal{J}\subset \mathcal{K}: j\in \mathcal{J}, \mathcal{B}(\bm{x},\mathcal{J}) = \mathcal{B}} \Prob\left(\mathcal{J} \text{ are receiving } \right)  \\
     & =\sum_{\mathcal{J}\subset \mathcal{K}: j\in \mathcal{J}, \mathcal{B}(\bm{x},\mathcal{J})= \mathcal{B}} \left(\prod_{l\in \mathcal{J}} \sigma_{l}\right) \left(\prod_{l\in \mathcal{K}\setminus \mathcal{J}}(1-\sigma_{l})\right) \label{eq:prob_active_UE_set}
    \end{align}
    where \eqref{eq:prob_active_UE_set} is due to the assumption that devices receive data independently.

    Since the service rate of each device depends on the set of APs having data, the service rate of device $j$ is chosen from $2^{n}$ possible values corresponding to different sets of APs having data. As in Sections~\ref{Spectrum allocation for fixed user association} and~\ref{Spectrum allocation with scheduling for fixed user association},
    we use a M/G/1 queue with different classes of packets to approximate the average delay of device $j$ in interactive queues.
    Each class corresponds to a specific set of interfering APs to device~$j$.
    In a stable queueing system, the arrival rate of packets for device $j$ when device in $\mathcal{B}$ have data is approximated as
    $\sigma_j p^{\mathcal{B}}_j r^\mathcal{B}_j$.
    The total arrival rate of device $j$ is
    \begin{align} \label{eq:arrival_rate_ue}
    \lambda_j = \sigma_j\sum_{\mathcal{B}\subset \mathcal{N}} p^{\mathcal{B}}_j r^\mathcal{B}_j.
    \end{align}
    According to formula (11) in \cite{Federgruen-1988}, the average delay of device $j$ using the M/G/1-queue approximation is
    \begin{align}
    \widehat{d}_j=  \frac{\sigma_j}{\lambda_j} + \frac{\sigma_j}{1-\sigma_j}\sum_{\mathcal{B}\subset \mathcal{N}: r^\mathcal{B}_j > 0} \frac{p^{\mathcal{B}}_j}{r^\mathcal{B}_j}.
    \end{align}


    With the preceding approximations, the joint dual-timescale resource allocation and user association problem is formulated as:
      \begin{problem}(Joint dual-timescale allocation and user association)  \label{Problem: UA_and_scheule_simplified}
      \begin{subequations}
       \begin{align}
        \underset{\bm{x}, \bm{y}, \bm{z},\bm{r}, \pmb{\sigma}, \bm{p}, \bm{\widehat{d}}}{\minimize} \ &  \sum_{j=1}^k \lambda_j  \widehat{d}_j   \tag{P6a} \\
        \subjectto
           &  \widehat{d}_j= \frac{\sigma_j}{\lambda_j} + \frac{\sigma_j}{1-\sigma_j}\sum_{\mathcal{B}\subset \mathcal{N}: r^\mathcal{B}_j > 0} \frac{p^{\mathcal{B}}_j}{r^\mathcal{B}_j}, \ \forall j\in \mathcal{K}  \tag{P6b} \\
           & r^\mathcal{B}_j= \frac{W}{L} \sum_{\mathcal{T}\subset \mathcal{N}}  z^\mathcal{T} \sum_{\mathcal{F}\subset \mathcal{N}} \left( \sum_{i\in \mathcal{N}}   \eta^{\mathcal{F},\mathcal{T},\mathcal{B}}_{i\rightarrow j} x_{i\rightarrow j}^\mathcal{F} \right)  , \ \forall j\in \mathcal{K}, \ \mathcal{B}\subset \mathcal{N} \tag{P6c} \\
          & \sum_{j\in \mathcal{K}}  x_{i\rightarrow j}^\mathcal{F}\leq y^\mathcal{F}, \ \forall i\in \mathcal{N}, \mathcal{F}\subset \mathcal{N}   \tag{P6d}\label{constr:p3_1}\\
          & \sum_{\mathcal{F}\subset \mathcal{N}} y^\mathcal{F} =1,  \tag{P6e} \\
          & \sum_{\mathcal{T}\subset \mathcal{N}}  z^\mathcal{T} =1,  \tag{P6f} \\
          &  z^\mathcal{T} \geq 0, \ \forall \mathcal{T}\subset \mathcal{N}  \tag{P6g} \\
          &  x_{i\rightarrow j}^\mathcal{F} \geq 0, \ \forall i\in \mathcal{N}, j\in \mathcal{K}, \mathcal{F}\subset \mathcal{N} \tag{P6h}  \\
          & 0\leq \sigma_j < 1, \ \forall j\in \mathcal{K} \tag{P6i} \\
          & \sigma_j p^{\mathcal{B}}_j  = \sum_{\mathcal{J}\subset \mathcal{K}: j\in \mathcal{J}, \mathcal{B}(\bm{x},\mathcal{J})= \mathcal{B}}  \left(\prod_{l\in \mathcal{J}} \sigma_{l}\right) \left(\prod_{l\in \mathcal{K}\setminus \mathcal{J}}(1-\sigma_{l})\right), \ \forall j\in \mathcal{K}, \ \mathcal{B}\subset \mathcal{N} \tag{P6j} \\
          & \lambda_j = \sigma_j\sum_{\mathcal{B}\subset \mathcal{N}} p^{\mathcal{B}}_j r^\mathcal{B}_j, \ \forall j\in \mathcal{K}. \tag{P6l}\label{constr:p3_3}
       \end{align}
      \end{subequations}
      \end{problem}

       The variables in Problem~\ref{Problem: UA_and_scheule_simplified} are $\bm{x},  \bm{y}, \bm{z}, \bm{r}=(r^\mathcal{B}_j)_{\mathcal{B}\subset \mathcal{N}, j\in \mathcal{K}},\pmb{\sigma}= (\sigma_j)_{j\in \mathcal{K}}$, $\bm{p}=(p^{\mathcal{B}}_j)_{\mathcal{B}\subset \mathcal{N}, j\in \mathcal{K}}$ and $\bm{\widehat{d}}= (\widehat{d}_j)_{j\in \mathcal{K}}$.
       In comparison with Problem~\ref{Problem:fixed_UA_with_scheule}, here we have new variables $\bm{x}$ and $\pmb{\sigma}$ as well as new constraints \eqref{constr:p3_1} and \eqref{constr:p3_3} (Recall that \eqref{constr:p3_1} is \eqref{eq:relate_x_y} and \eqref{constr:p3_3} is \eqref{eq:arrival_rate_ue}).
       Problem~\ref{Problem: UA_and_scheule_simplified} may be nonconvex due to the nonlinear-equality constraints. Similar to in the previous sections, we can use iterative method to solve the problem with low complexity. We divide Problem~\ref{Problem: UA_and_scheule_simplified} into three sub-problems: The first one is to update $\bm{x}$,  $\bm{y}$ and $\bm{r}$ by fixing $\pmb{\sigma}$, $\bm{p}$ and $\bm{z}$; the second one is to update $\bm{z}$ and $\bm{r}$ by fixing $\pmb{\sigma}$, $\bm{p}$, $\bm{x}$ and $\bm{y}$; the third one is to update $\pmb{\sigma}$ and $\bm{p}$ by fixing $\bm{x},\bm{y}, \bm{z}$ and $\bm{r}$.

    \begin{problem}(For fixed $\pmb{\sigma}$, $\bm{p}$ and $\bm{z}$, update $\bm{x}$,  $\bm{y}$ and $\bm{r}$) \label{Problem:UA_scheule_simplified_sub1}
      \begin{subequations}
       \begin{align}
         \underset{\bm{x}, \bm{y}, \bm{r}}{\minimize} \ &  \sum_{j\in \mathcal{K}} \sigma_j +
           \sum_{j\in \mathcal{K}} \sum_{\mathcal{B}\subset \mathcal{N}: r^\mathcal{B}_j>0} \frac{\sigma_j p^{\mathcal{B}}_j\lambda_j}{r^\mathcal{B}_j(1-\sigma_j)} \tag{P7a} \\
        \subjectto
           & r^\mathcal{B}_j=\frac{W}{L} \sum_{\mathcal{T}\subset \mathcal{N}}  z^\mathcal{T} \sum_{\mathcal{F}\subset \mathcal{N}} \left( \sum_{i\in \mathcal{N}}   \eta^{\mathcal{F},\mathcal{T},\mathcal{B}}_{i\rightarrow j} x_{i\rightarrow j}^\mathcal{F} \right), \ \forall j\in \mathcal{K}, \ \mathcal{B}\subset \mathcal{N}  \tag{P7b}\label{eq:r_constraint_UE_I_0}\\
          & \sum_{j\in \mathcal{K}}  x_{i\rightarrow j}^\mathcal{F}\leq y^\mathcal{F}, \ \forall i\in \mathcal{N}, \mathcal{F}\subset \mathcal{N}  \tag{P7c} \\
          & \sum_{\mathcal{F}\subset \mathcal{N}} y^\mathcal{F} =1, \tag{P7d}  \\
          &  x_{i\rightarrow j}^\mathcal{F} \geq 0, \ \forall i\in \mathcal{N}, j\in \mathcal{K}, \mathcal{F}\subset \mathcal{N}  \tag{P7e} \\
          & \lambda_j = \sigma_j\sum_{\mathcal{B}\subset \mathcal{N}} p^{\mathcal{B}}_j r^\mathcal{B}_j, \ \forall j\in \mathcal{K}.  \tag{P7f}\label{ineq:decreasing_rho_3}
       \end{align}
      \end{subequations}
    \end{problem}

     \begin{problem}(For fixed $\pmb{\sigma}$, $\bm{p}$, $\bm{x}$ and $\bm{y}$, update $\bm{z}$ and $\bm{r}$) \label{Problem:UA_scheule_simplified_sub2}
      \begin{subequations}
       \begin{align}
        \underset{\bm{x}, \bm{y}, \bm{r}}{\minimize} \ &  \sum_{j\in \mathcal{K}} \sigma_j +
           \sum_{j\in \mathcal{K}} \sum_{\mathcal{B}\subset \mathcal{N}: r^\mathcal{B}_j>0} \frac{\sigma_j p^{\mathcal{B}}_j\lambda_j}{r^\mathcal{B}_j(1-\sigma_j)} \tag{P8a} \\
        \subjectto
           & r^\mathcal{B}_j= \frac{W}{L} \sum_{\mathcal{T}\subset \mathcal{N}}  z^\mathcal{T} \sum_{\mathcal{F}\subset \mathcal{N}} \left( \sum_{i\in \mathcal{N}}   \eta^{\mathcal{F},\mathcal{T},\mathcal{B}}_{i\rightarrow j} x_{i\rightarrow j}^\mathcal{F} \right)  , \ \forall j\in \mathcal{K}, \ \mathcal{B}\subset \mathcal{N}  \tag{P8b}\label{eq:r_constraint_UE_I}\\
          & \sum_{\mathcal{T}\subset \mathcal{N}}  z^\mathcal{T} =1,  \tag{P8c} \\
          &  z^\mathcal{T} \geq 0, \ \forall \mathcal{T}\subset \mathcal{N}   \tag{P8d} \\
          & \lambda_j = \sigma_j\sum_{\mathcal{B}\subset \mathcal{N}} p^{\mathcal{B}}_j r^\mathcal{B}_j, \ \forall j\in \mathcal{K}. \tag{P8e}\label{ineq:decreasing_rho_4}
       \end{align}
      \end{subequations}
      \end{problem}

      Problems~\ref{Problem:UA_scheule_simplified_sub1} and~\ref{Problem:UA_scheule_simplified_sub2} are convex optimization problems.
      Denote $\bm{g}(\pmb{\sigma})$ as a vector of $k$ functions, where the $j$th function is defined as
    \begin{align}  \label{def:interf_func_ue}
    g_j(\pmb{\sigma})=\frac{\lambda_j}{ \sum_{\mathcal{J}\subset \mathcal{K}: j\in \mathcal{J}} \left(\prod_{l\in \mathcal{J}\setminus \{j\}} \sigma_l\right) \left(\prod_{l\in \mathcal{K}\setminus \mathcal{J}}(1-\sigma_l)\right)  r^{\mathcal{B}(\bm{x},\mathcal{J})}_j}.
    \end{align}
    By substituting \eqref{eq:prob_active_UE_set} into \eqref{eq:arrival_rate_ue}, we eliminate $(p^{\mathcal{B}}_j)_{\mathcal{B}\subset \mathcal{N}, j\in \mathcal{K}}$ and obtain $\sigma_j = g_j(\pmb{\sigma})$, so $\pmb{\sigma}$ is a fixed point of
    \begin{align} \label{eq:sigma}
    \pmb{\sigma} = \bm{g}(\pmb{\sigma}).
    \end{align}
        We also explore the properties of function $\bm{g}(\cdot)$ as Lemma~\ref{lemma:properties}.
    \begin{lemma} \label{lemma:properties_Hhat}
    Let $(\lambda_j)_{j\in \mathcal{K}}$ and $(r^\mathcal{B}_j)_{\mathcal{B}\subset \mathcal{N}, j\in \mathcal{K}}$ be fixed.
    If $0\leq \pmb{\widetilde{\sigma}}\leq \pmb{\sigma}< 1$, then $\bm{g}(\pmb{\sigma}) \geq \bm{g}(\pmb{\widetilde{\sigma}})$.
    \end{lemma}
    The proof of Lemma \ref{lemma:properties_Hhat} is similar to Lemmas \ref{lemma:properties} and is omitted here. Given $\bm{x}$, $\bm{y}$, $\bm{z}$ and $\bm{r}$, we update $\pmb{\sigma}$ and $\bm{p}$ through an iterative method similar to Algorithm~\ref{alg:update_rho}:
    \begin{algorithm}(For fixed $\bm{x},\bm{y}, \bm{z}$ and $\bm{r}$, update $\pmb{\sigma}$ and $\bm{p}$): \label{alg:update_sigma_rho}
    \begin{itemize}
        \item[1.] Initialization: set $\pmb{\sigma}^{(0)}$ as the current device utilizations and let $\epsilon$ be a small positive constant;
        \item[2.] Update utilization and probabilities:
        Repeat
            \begin{align}
                & \pmb{\sigma}^{(m+1)} = \bm{g}(\pmb{\sigma}^{(m)}) \label{eq:alg_sigma_1}
            \end{align}
        until $|\pmb{\sigma}^{(m)}-\pmb{\sigma}^{(m-1)}| < \epsilon$;
        \item[3.] Let
        \begin{align}
                & \sigma_j = \sigma_j^{(m)}, \ \forall j\in \mathcal{K}
        \end{align}
         and compute $(p^{\mathcal{B}}_j)_{\mathcal{B}\subset \mathcal{N}, j\in \mathcal{K}}$ by \eqref{eq:prob_active_UE_set}.
     \end{itemize}
    \end{algorithm}
    Then we demonstrate the convergence of Algorithm~\ref{alg:update_sigma_rho}.

    \begin{lemma} \label{lemma:alg_sigma_convg}
    Algorithm~\ref{alg:update_sigma_rho} converges when it starts with $\pmb{\sigma}^{(0)}$ satisfying $\pmb{\sigma}^{(0)}\geq \bm{g}(\pmb{\sigma}^{(0)})$.
    \end{lemma}
    The proof of Lemma \ref{lemma:alg_sigma_convg} is similar to Lemma \ref{lemma:alg_rho_convg} and is omitted here. 
    After introducing each problem in the proposed method, we illustrate an iterative algorithm to solve Problem~\ref{Problem: UA_and_scheule_simplified}:
    \begin{algorithm}(Iterative algorithm for solving Problem~\ref{Problem: UA_and_scheule_simplified}): \label{Alg:UA_scheule_simplified}
    \begin{itemize}
    \item[1.] Initialization: set $\sigma_j=1$ for all $j\in \mathcal{K}$; $p^{\mathcal{B}}_j=1$ if $\mathcal{B}=\mathcal{N}$ and $j\in \mathcal{K}$, $p^{\mathcal{B}}_j=0$ otherwise;
    \item[2.] Update $\bm{x}$,  $\bm{y}$ and $\bm{r}$ by solving Problem~\ref{Problem:UA_scheule_simplified_sub1};


    \item[3.] Update $\bm{z}$ and $\bm{r}$ by solving Problem~\ref{Problem:UA_scheule_simplified_sub2};

    \item[4.] Update $\pmb{\sigma}$ and $\bm{p}$ using Algorithm~\ref{alg:update_sigma_rho};

    \item[5.] Terminate if $\bm{x}$ has converged. Otherwise, return to step 2.

    \end{itemize}
    \end{algorithm}
    Using Algorithm \ref{Alg:UA_scheule_simplified}, Problem~\ref{Problem: UA_and_scheule_simplified} can be solved in an iterative way with manageable complexity.

    \begin{theorem}  \label{Thm3:convergence}
     Algorithm~\ref{Alg:UA_scheule_simplified} converges to a fixed point.
    \end{theorem}
    The proof is similar to that for Theorem~\ref{Thm1:convergence} and is omitted here.



    \section{Simulation Results} \label{Simulation results}

\begin{figure}
    \centering
    \includegraphics[angle=360,width=.6\columnwidth]{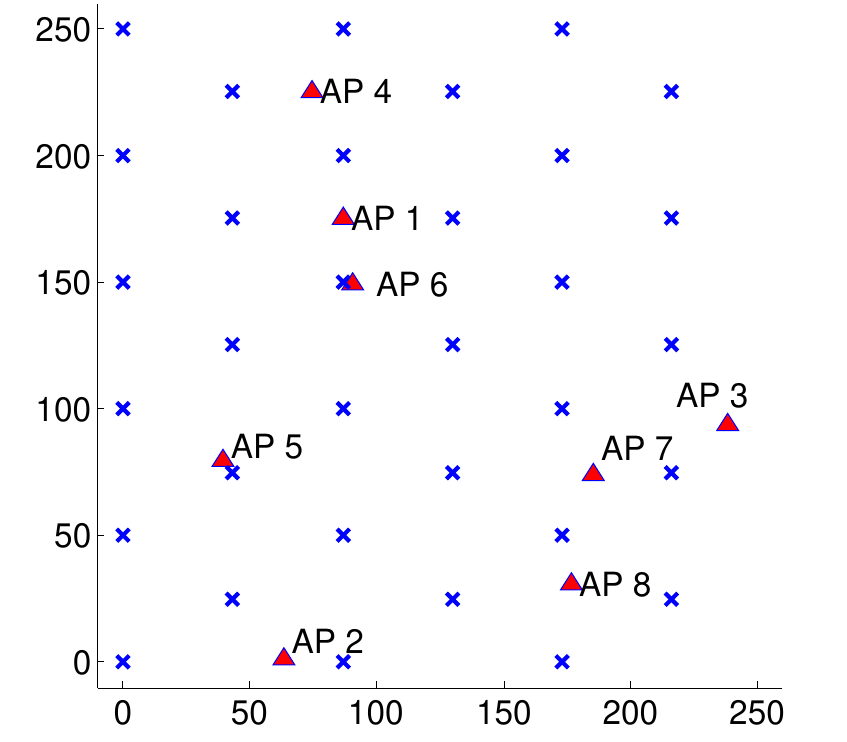}
    \caption{A system with 8 APs and 33 devices (represented by '$\times$').}
    \label{fig:Asiloma_all_top_8APs}
    \end{figure}
\begin{figure}
    \centering
    \includegraphics[width=0.6\columnwidth]{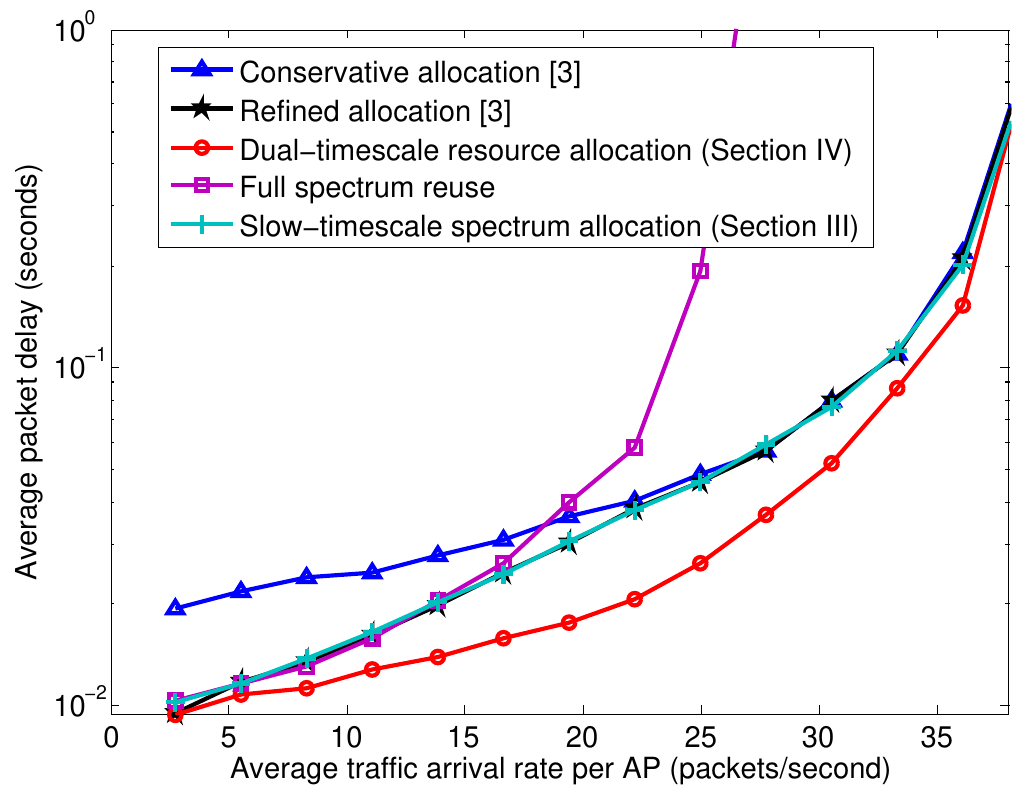}
    \caption{Delay performance of different allocation methods with fixed user association as a
    function of average network traffic load.}
    \label{fig:sim results optimization problems for fixed user association}
    \end{figure}

    First we show the performance of the proposed resource allocation for fixed user association (described in Section~\ref{Spectrum allocation for fixed user association} and~\ref{Spectrum allocation with scheduling for fixed user association}).
    A system with eight APs in the region of 250 meter by 250 meter is studied, where APs are located as in Fig.~\ref{fig:Asiloma_all_top_8APs},
    but each AP is associated with one device (only the corresponding closest devices to the eight APs in Fig.~\ref{fig:Asiloma_all_top_8APs} are included here, the remaining ones are used in a later simulation).
    The proposed resource allocations (moderate-timescale allocations, and dual-timescale resource allocation) are compared
    with three other schemes: 1) full-spectrum reuse (as in LTE),
    2) the conservative allocation in~\cite{ZB-2014}, and
    3) the 
    refined 
    allocation 
    in \cite{ZB-2014}.
    In the experiments, interactive queues are simulated where service rates adapt to instantaneous interference.
    Fig.~\ref{fig:sim results optimization problems for fixed user association} depicts average packet delay against different traffic loads for all four different 
    allocations.
    As expected, the proposed dual-timescale allocation (with opportunistic fast-timescale scheduling) has the best performance under all traffic conditions. The refined allocation and moderate-timescale allocation using the M/G/1-queue approximation are the close second in both cases of very light traffic and very heavy traffic, but leads to much higher delay in case of moderate traffic. Indeed, this case is when it is the most beneficial to consider the impact of fast-timescale scheduling when making moderate-timescale allocations.

    \begin{figure}
    \centering
    \includegraphics[angle=360,width=.6\columnwidth]{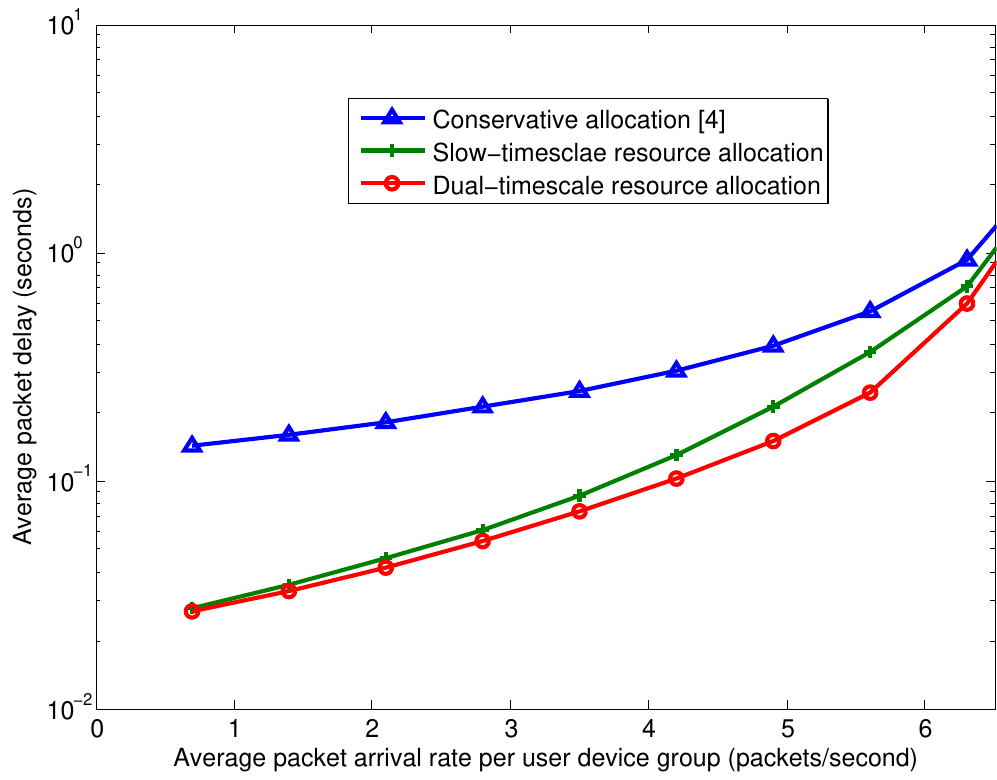}
    \caption{Delay performance of joint resource allocation and user association.}
    \label{fig:Asiloma_all_sim_8APs}
    \end{figure}

    Next, we show the performance of the joint resource allocation and user association.
    We consider a system with 8 APs and 33 devices in Fig.~\ref{fig:Asiloma_all_top_8APs}.
    Resource allocation by the proposed algorithm is compared with the conservative model in \cite{ZB-2018}. Interactive queues are simulated by using the solution to the proposed optimization problems.
    Simulation results are illustrated in Fig.~\ref{fig:Asiloma_all_sim_8APs}.
    Resource allocation on the moderate timescale, which is a special case of the formulation in Section \ref{Spectrum allocation with scheduling for flexible user association} with $z^\mathcal{N} =1$, has marginal gain over the conservative model. With dual-timescale allocation, the average delay is further reduced. As the average packet arrival rate increases, the gain by dual-timescale allocation goes up first, then reduces. When the average arrival rate per device is 5.5 packets per second, we observe about $30\%$ gain for dual-timescale resource allocation over moderate-timescale resource allocation. Based on observations in the experiments, the gain by dual-timescale allocation is substantial in the interference-limited case. It is because that APs can take advantage of opportunistic scheduling on a fast timescale and achieve much higher rate when their neighbors are silent.

%

%

%
    \section{Conclusion} \label{Conclusion}


    We have reported some new modeling, problem formulation, and techniques for traffic-driven radio resource management in wireless heterogeneous networks.  At the core is a global moderate-timescale optimization with considerations of effects of dynamic scheduling on the fast timescale as well as utilization of allocated resources.  Exploiting these factors leads to substantial quality of service improvements. The improvements are obtained at the cost of increased computational complexity. The scheme is readily implemented in a relatively small network.  Extension of the work to large networks is left to future work.

\appendices

\section{Proof of Lemma \ref{lemma:MG1_delay}} \label{appdx:MG1_delay}

We prove this lemma using the delay analysis for M/G/c queue with multiple customer classes in \cite{Federgruen-1988}. According to formula (11) in \cite{Federgruen-1988}, the approximation of average waiting time for the general case is
\begin{align} \label{eq:MG1_delay}
w = \frac{\left(\Exp [V^2]\right) b}{2\Exp[V](c-\rho)},
\end{align}
where $\lambda$ is total arrival rate, $V$ is the service time random variable, $\rho = \lambda \Exp [V]$, $c$ is the number of servers, and $b$ is the probability that all servers are occupied in an M/M/c system with the same expected service time.
The approximation \eqref{eq:MG1_delay} becomes exact in our case of a single server (c=1) \cite{Federgruen-1988}.

Consider the M/G/1 queue at AP $i$. According to Little's Theorem \cite{Bertsekas-Gallager-1987}, the probability that the server is occupied in the M/M/1 queue with the same expected service time $\Exp[V]$ is
\begin{align} \label{eq:B}
b =\lambda_i \Exp[V].
\end{align}
Hence, from \eqref{eq:MG1_delay} and \eqref{eq:B}, the average waiting time for the M/G/1 queue is
\begin{align} \label{eq:MG1_delay_2}
w = \frac{\left(\Exp [V^2]\right)\lambda_i}{2(1-\rho_i)}.
\end{align}

According to our formulation, AP $i$ has $2^{n-1}$ classes of packets. The service rate of each class takes a value from the set $\left(r_i^\mathcal{B}\right)_{i\in \mathcal{B}\subset \mathcal{N}}$. The utilization of the packets corresponding to service rate $r_i^\mathcal{B}$ is $p^\mathcal{B}$. Thus, the arrival rate of such packets $\lambda_i^{\mathcal{B}}$ is equal to  $p^\mathcal{B}r^\mathcal{B}_i$, and the total utilization at AP $i$ is equal to
\begin{align} \label{eq:rho_MG1}
       \sum_{\mathcal{B}\subset \mathcal{N}: i\in \mathcal{B}} p^\mathcal{B}  = \rho_i.
    \end{align}
Given that the packet length of each class follows an exponential distribution, the expected service time of each packet with service rate $r^\mathcal{B}_i$ is $\frac1{r^\mathcal{B}_i}$, and the variance of the service time is also $\frac1{r^\mathcal{B}_i}$. The proportion of the packets with service rate $r^\mathcal{B}_i$ is $\frac{\lambda_i^\mathcal{B}}{\lambda_i }$, thus the expected service time of each packet among all classes is
\begin{align}
    \Exp[V] & = \sum_{\mathcal{B}\subset \mathcal{N}: i\in \mathcal{B}} \frac{\lambda_i^\mathcal{B}}{\lambda_i } \cdot \frac1{r^\mathcal{B}_i} \\
     & = \sum_{\mathcal{B}\subset \mathcal{N}: i\in \mathcal{B}} \frac{p^\mathcal{B}r^\mathcal{B}_i}{\lambda_i } \cdot \frac1{r^\mathcal{B}_i} \\
    & = \frac{\rho_i}{\lambda_i}, \label{eq:expected_service_time}
\end{align}
and the second moment of the service time is
\begin{align}
    \Exp[V^2] & = \sum_{\mathcal{B}\subset \mathcal{N}: i\in \mathcal{B}} \frac{\lambda_\mathcal{B}^i}{\lambda_i } \cdot \left(\frac{1}{\left(r^\mathcal{B}_i\right)^2} + \frac{1}{\left(r^\mathcal{B}_i\right)^2}\right) \\
    & = \sum_{\mathcal{B}\subset \mathcal{N}: i\in \mathcal{B}} \frac{p^\mathcal{B}r^\mathcal{B}_i}{\lambda_i } \cdot \frac{2}{\left(r^\mathcal{B}_i\right)^2}  \\
     & = \frac{2}{\lambda_i}\sum_{\mathcal{B}\subset \mathcal{N}: i\in \mathcal{B}} \frac{p^\mathcal{B}}{ r^\mathcal{B}_i}. \label{eq:exp_v_sq}
\end{align}

From \eqref{eq:MG1_delay_2} and \eqref{eq:exp_v_sq}, the average service time of AP $i$ is
\begin{align}
     w  & = \frac1{1-\rho_i}\cdot \sum_{\mathcal{B}\subset \mathcal{N}: i\in \mathcal{B}} \frac{p^\mathcal{B}}{ r^\mathcal{B}_i}. \label{AppdxA_service_time}
\end{align}
Therefore, from \eqref{eq:expected_service_time} and \eqref{AppdxA_service_time}, the average delay of AP $i$ under our formulation is
\begin{align}
     d_i & = \Exp[V] + w  \\
         & = \frac{\rho_i}{\lambda_i} + \frac1{1-\rho_i}\cdot \sum_{\mathcal{B}\subset \mathcal{N}: i\in \mathcal{B}} \frac{p^\mathcal{B}}{ r^\mathcal{B}_i}.
\end{align}

    \section{Proof of Lemma \ref{lemma:properties}} \label{Appdx:lemma2}
     If $\pmb{\rho}\geq \pmb{\widehat{\rho}}$, then $\rho_l\geq \widehat{\rho}_l$ for all $l\in \mathcal{N}$. To prove this lemma, it is sufficient to show that for all $i\in \mathcal{N}$, $f_i(\pmb{\rho})$ is non-decreasing in each element of $\pmb{\rho}$.

     According to \eqref{def:interf_func_0}, for all $i,j\in \mathcal{N}$ we have
     \begin{align}
     \frac{\partial g_i(\pmb{\rho})}{\partial \rho_j} & = \frac{\partial }{\partial \rho_j} \left(\sum_{\mathcal{B}\subset \mathcal{N}: i\in \mathcal{B}} \prod_{l\in \mathcal{B}\setminus \{i\}} \rho_l \prod_{l\in \mathcal{N}\setminus \mathcal{B}}(1-\rho_l) r^\mathcal{B}_i\right) \label{eq:L2_1} \\
     & = \frac{\partial }{\partial \rho_j} \left(\sum_{\mathcal{B}\subset \mathcal{N}: i, j\in \mathcal{B}} \left(\prod_{l\in \mathcal{B}\setminus \{i\}} \rho_l \prod_{l\in \mathcal{N}\setminus \mathcal{B}}(1-\rho_l)  r^\mathcal{B}_i  +
     \prod_{l\in \mathcal{B}\setminus \{i,j\}} \rho_l \prod_{l\in (\mathcal{N}\setminus \mathcal{B})\bigcup \{j\}}(1-\rho_l)  r^{\mathcal{B}\setminus \{j\}}_i
     \right) \right) \label{eq:L2_2} \\
     & = \sum_{\mathcal{B}\subset \mathcal{N}: i, j\in \mathcal{B}} \left(\prod_{l\in \mathcal{B}\setminus \{i,j\}} \rho_l \prod_{l\in \mathcal{N}\setminus \mathcal{B}}(1-\rho_l)  r^\mathcal{B}_i  -
     \prod_{l\in \mathcal{B}\setminus \{i,j\}} \rho_l \prod_{l\in \mathcal{N}\setminus \mathcal{B}}(1-\rho_l)  r^{\mathcal{B}\setminus \{j\}}_i
     \right)  \label{eq:L2_3} \\
     & = \sum_{\mathcal{B}\subset \mathcal{N}: i, j\in \mathcal{B}} \left(\prod_{l\in \mathcal{B}\setminus \{i,j\}} \rho_l \prod_{l\in \mathcal{N}\setminus \mathcal{B}}(1-\rho_l)  \left(r^\mathcal{B}_i  -  r^{\mathcal{B}\setminus \{j\}}_i\right)
     \right)  \label{eq:L2_4} \\
     & \leq  0,  \label{eq:L2_5}
    \end{align}
    where we use the fact that $r^\mathcal{B}_i \leq r^{\mathcal{B}\setminus \{j\}}_i$, which is due to \eqref{eq:SE_comparison}.

    From \eqref{def:interf_func} and \eqref{eq:L2_5}, we have
     \begin{align}
    \frac{\partial f_i(\pmb{\rho})}{\partial \rho_j} & = -\frac{\lambda_i}{(g_i(\pmb{\rho}))^2}\frac{\partial g_i(\pmb{\rho})}{\partial \rho_j} \\
    & \geq 0.
    \end{align}
    Hence, $f_i(\pmb{\rho})$ is non-decreasing in each element of $\pmb{\rho}$.

    \section{Proof of Lemma \ref{lemma:alg_rho_convg}} \label{appdx:lemma4}

    From Algorithm~\ref{alg:update_rho}, $\pmb{\rho}^{(1)} = \bm{f}(\pmb{\rho}^{(0)})$. If $\pmb{\rho}^{(0)}\geq \bm{f}(\pmb{\rho}^{(0)})$, then according to Lemma~\ref{lemma:properties}
    \begin{align}
    \pmb{\rho}^{(1)} =\bm{f}(\pmb{\rho}^{(0)}) \geq \bm{f}(\bm{f}(\pmb{\rho}^{(0)}))= \bm{f}(\pmb{\rho}^{(1)}) =\pmb{\rho}^{(2)}.
    \end{align}
    By induction, for any $m=0,1,2,\cdots$,
    \begin{align}
    \pmb{\rho}^{(m+1)}\leq \pmb{\rho}^{(m)}.
    \end{align}
    Because series $(\pmb{\rho}^{(m)})_{m\in \mathcal{N}}$ is lower-bounded by zero and non-increasing, the series must converge to the fixed point of $\pmb{\rho} =\bm{f}(\pmb{\rho})$.

    \section{Proof of Theorem \ref{Thm1:convergence}} \label{Appdx:thm1}

    First, we prove Problem~\ref{Problem:fixed_UA_no_scheule_sub1} in the step 2 of Algorithm~\ref{Alg:fixed_UA_no_scheule} is always feasible for any $\pmb{\lambda}\in \Lambda$.  Here an iteration is referred as a combination of steps 2 and 3. In the first iteration, we have $\pmb{\rho}=\mathbf{1}$ in step 2.
    It is easy to check that Problem~\ref{Problem:fixed_UA_no_scheule_sub1} is feasible for $\pmb{\rho}=\mathbf{1}$ as long as $\pmb{\lambda}\in \Lambda$ according to the results in \cite{Tassiulas-1922}.

    We show that if Problem~\ref{Problem:fixed_UA_no_scheule_sub1} in the $m$th iteration is feasible, then it is also feasible in the $(m+1)$st iteration.
    Assuming that Problem~\ref{Problem:fixed_UA_no_scheule_sub1} is feasible given $\pmb{\rho}$ and $\bm{p}$ in step 2 of the $m$th iteration, the solution to Problem~\ref{Problem:fixed_UA_no_scheule_sub1} $(\bm{y},\bm{r})$ satisfies \eqref{ineq:decreasing_rho}. From \eqref{eq:prob_active_set}, \eqref{ineq:decreasing_rho} and \eqref{def:interf_func}, $\bm{y}$ and $\bm{r}$ in step 2 result in $\bm{f}(\pmb{\rho}) \leq \pmb{\rho}$. Consequently, at the beginning of step 3 of the $m$th iteration,
    we define $\pmb{\rho}^{(0)}  = \pmb{\rho}$. Since $\bm{f}(\pmb{\rho}^{(0)}) = \bm{f}(\pmb{\rho}) \leq \pmb{\rho} = \pmb{\rho}^{(0)}$, according to Lemma~\ref{lemma:alg_rho_convg}, Algorithm~\ref{alg:update_rho} in the step 3 of the $m$th iteration converges. The result of Algorithm~\ref{alg:update_rho} $(\pmb{\widehat{\rho}},\bm{\widehat{p}})$ satisfies \eqref{eq:prob_active_set} and \eqref{eq:rho}. In addition, $\pmb{\widehat{\rho}}$ is smaller that $\pmb{\rho}^{(0)}$.
    Next, the algorithm goes to step 2 of the $(m+1)$st iteration. With $(\bm{y}, \bm{r})$ from step 2 of the $m$th iteration and $(\pmb{\widehat{\rho}},\bm{\widehat{p}})$ from step 3 of the $m$th iteration, \eqref{ineq:decreasing_rho} holds. Thus, $(\bm{y}, \bm{r})$ satisfies all the constraints in Problem~\ref{Problem:fixed_UA_no_scheule_sub1} for fixed $(\pmb{\widehat{\rho}},\bm{\widehat{p}})$. Therefore, Problem~\ref{Problem:fixed_UA_no_scheule_sub1} is feasible in step 2 of the $(m+1)$st iteration .

    Furthermore, from the preceding analysis, an iteration consisting of steps 2 and 3 results in non-increasing $\pmb{\rho}$ which is lower-bounded by zero. Hence, Algorithm~\ref{Alg:fixed_UA_no_scheule} converges to a fixed point.

\bibliography{references}{}
\bibliographystyle{ieeetr}

%
%

%

\end{document}